\documentclass[12pt,superscriptaddress]{iopart}
\usepackage[T1]{fontenc}
\usepackage{color}
\usepackage{graphicx,amsfonts,amsbsy,amssymb,amsthm}
\usepackage{epstopdf}
\usepackage{dcolumn}
\usepackage{braket}
\usepackage{hyperref}

\def\mc {\mathcal}

\def\diag {{\mathrm{diag}}}

\begin{document}
\title[Repulsively diverging gradient of the density functional]{Repulsively diverging gradient of the density functional in the Reduced Density Matrix Functional Theory}

\author{Tomasz Maci\k{a}\.{z}ek}
\ead{tomasz.maciazek@bristol.ac.uk}
\address{School of Mathematics, University of Bristol, Fry Building, Woodland Road, Bristol BS8 1UG, UK}

\begin{abstract}
The Reduced Density Matrix Functional Theory (RDMFT) is a remarkable tool for studying properties of ground states of strongly interacting quantum many body systems. As it gives access to the one-particle reduced density matrix of the ground state, it provides a perfectly tailored approach to studying the Bose-Einstein condensation or systems of strongly correlated electrons. In particular, for homogeneous Bose-Einstein condensates as well as for the Bose-Hubbard dimer it has been recently shown that the relevant density functional exhibits a repulsive gradient (called the Bose-Einstein condensation force) which diverges when the fraction of non-condensed bosons tends to zero. In this paper, we show that the existence of the Bose-Einstein condensation force is completely universal for any type of pair-interaction and also in the non-homogeneous gases. To this end, we construct a universal family of variational trial states which allows us to suitably approximate the relevant density functional in a finite region around the set of the completely condensed states. We also show the existence of an analogous repulsive gradient in the fermionic RDMFT for the $N$-fermion singlet sector in the vicinity of the set of the Hartree-Fock states. Finally, we show that our approximate functional may perform well in electron transfer calculations involving low numbers of electrons. This is demonstrated numerically in the Fermi-Hubbard model in the strongly correlated limit where some other approximate functionals are known to fail.
\end{abstract}

\maketitle
\section{Introduction}
The Bose-Einstein condensate (BEC) has been theoretically predicted by Bose and Einstein \cite{einstein,bose} in 1924. Seventy years passed before it was finally realised experimentally by Wieman and Cornell and independently by Ketterle in gases of ultracold atoms \cite{cornell,ketterle}. Nowadays, the BEC provides a widely-used experimental testing ground for quantum many-body theories, especially superfluidity, superconductivity \cite{bcs1,bcs2,bcs3} and the laser phenomenon \cite{laser,review}. In this work, we focus on the description of the BEC via the Density Functional Theories. Our developed methods for the BEC are subsequently extended to systems of strongly interacting electrons in the singlet sector.


The most successful and renowned functional theory is perhaps the one which uses the Hohenberg-Kohn density functional \cite{HK} (see also \cite{lieb}). The natural variable of the Hohenberg-Kohn density functional is the single-particle density which unfortunately is not suitable for the purpose of this paper \cite{penrose}. One of the reasons is that the single-particle densities of the fully condensed quantum states can range through all possible densities. In other words, it is not possible to decide whether a given single-particle density describes a BEC condensate or not. Thus, one needs to use a more sophisticated generalization of the Hohenberg-Kohn density functional, namely the Reduced Density Matrix Functional Theory (RDMFT) due to Gilbert \cite{Gilbert} and Levy \cite{levy}. RDMFT uses the single-particle reduced density matrix as its natural variable and hence is able to recover quantum correlations exactly. All density functional theories suffer the issue that their respective functionals are extremely challenging to find and hence they are usually not known. Remarkable exceptional cases in RDMFT where the functional is actually known analytically include the Bose-Hubbard dimer occupied by two bosons \cite{schilling-bosons} and the singlet sector of the Fermi-Hubbard dimer for two spinful electrons \cite{cohen1,cohen2}. The common practical strategy is thus to find suitable approximations to the relevant density functional. In RDMFT it is often the case that approximate functionals for some specific systems often lead to progress in finding more complicated approximate functionals for more general systems. For instance, in the fermionic RDMFT this has been the case with the well-known Hartree-Fock functional \cite{lieb-HF} which in turn led to the discovery of the M\"{u}ller and Power functionals \cite{muller,power} which in turn inspired numerous more complicated hierarchies of functionals \cite{pernal,challenges}. It is also a crucial task to construct approximate functionals that are simple enough to make explicit calculations tractable, but accurate enough to capture some desired information about the exact functional. Our presented work makes effort in this direction. Our results build on the recent tremendous progress in the foundations of both bosonic and fermionic RDMFT \cite{schilling-search,schilling-force,schilling-bosons,schilling-bec}, especially in RDMFT for bosons and homogeneous Bose-Einstein condensates, where the remarkable concept of the repulsive {\it BEC force} has been introduced. In particular, it has been shown that the gradient of the relevant density functional in the space of the reduced density matrices diverges repulsively in the regime of the Bose-Einstein condensation as the inverse of the square root of the fraction of the non-condensed bosons. This provided an alternative and more fundamental explanation for the existence of the quantum depletion in the homogeneous condensate \cite{schilling-bec} (see also \cite{girardeau,seiringer1, seiringer2} for the underlying theory) and in the Bose-Hubbard dimer \cite{schilling-bosons}.  This is compared with the celebrated theories by Bogoliubov \cite{bogoliubov} and Gross–Pitaevskii \cite{pitaevskii} which work in the dilute or weakly interacting regime as well as in the the high density regime for charged bosons and that have been recently verified experimentally \cite{lopes}.

Our work shows that the repulsive BEC force is present also in non-homogeneous Bose-Einstein condensates, showing that the presence of the BEC force is an universal phenomenon making the above alternative explanation of the quantum depletion complete for any system of pairwise interacting bosons. Our arguments are remarkably versatile and they can be extended to other physical systems. In particular, similar methods are employed to prove the existence of an analogous phenomenon for the singlet fermionic systems. Our construction coincides with the exact functional in the dimer case (two fermions occupying two sites/orbitals) known from previous works \cite{cohen1}. We numerically test the accuracy of our approximate functional in the Fermi-Hubbard model. For two fermions ($N=2$) we show that the approximate functional gives very accurate values of the ground state energy. For the Fermi-Hubbard chains, we employ the approximate functional for calculating the electron transfer. In particular, for the half-filled chain of length four, our functional recovers the exact electron transfer in the strongly correlated limit with a good accuracy. Remarkably, this is in contrast with the M\"{u}ller and Power functionals \cite{muller,power} which are known to fail even in the dimer case \cite{cohen1}. This shows that our proposed functional or its generalizations (see Section \ref{sec:discussion}) may be useful in electron transfer calculations, a task which has been pointed out as one of the current challenges of the RDMFT \cite{transfer,challenges}.

For the sake of clarity, we would like to make precise what is the form of the most general bosonic hamiltonian with pairwise interactions that we consider. We work in the discrete setting and consider quantum systems of $N$ bosons occupying $d$ sites interacting according to hamiltonians of the form
\begin{equation}\label{hamiltonian}
\hat H=\hat h +\hat W,
\end{equation}
where $\hat h$ is the single-particle term which comprises of hopping and local potentials while $\hat W$ is an arbitrary pair interaction. Thus, the most general form of $\hat h$ is
\[\hat h= \sum_{i<j}^d\left(t_{ij}b_i b_j^\dagger +\overline{t_{ij}}b_jb_i^\dagger \right)+\sum_{i=1}^dv_i\hat n_i,\]
where $t_{i,j}$ are (possibly complex) hopping amplitudes and $v_i$ are on-site potentials. Recall that the creation and annihilation operators $b_i$, $b_j^\dagger$ satisfy the bosonic commutation rules $[b_i,b_j^\dagger]=\delta_{i,j}$, $[b_i^\dagger,b_j^\dagger]=[b_i,b_j]=0$ and $\hat n_i:= b_i^\dagger b_i$ is the occupation number operator associated with site $i$. Similarly, the most general form of the pair interaction reads
\begin{equation}\label{interaction-general}
\hat W=\sum_{i,j,k,l=1}^d \Omega_{ijkl}b_ib_jb_k^\dagger b_l^\dagger +h.c.,
\end{equation}
where $\Omega_{ijkl}$ are complex parameters.

\section{A brief recap of the RDMFT}
The setting of the Reduced Density Matrix Functional Theory is to fix the interaction term $\hat W$ in \Eref{hamiltonian} and consider a family of hamiltonians $\hat H(h)=\hat h+\hat W$. One of the key objects in RDMFT is the one-particle reduced density matrix (denoted here shortly by 1RDM) which in the case of $N$ bosons occupying $d$ sites is a $d\times d$ hermitian matrix assigned to any state $\Ket{\Psi}$ whose entries read
\begin{equation}\label{1rdm-def}
\gamma\left(\Ket{\Psi}\right)_{i,j}=\Bra{\Psi}b_jb_i^\dagger \Ket{\Psi}.
\end{equation}
For any one particle hamiltonian $\hat h=\sum_{i,j}h_{i,j}b_i b_j^\dagger$ we have
\[\Bra{\Psi}\hat h \Ket{\Psi}=\sum_{i,j}h_{i,j}\gamma\left(\Ket{\Psi}\right)_{j,i}=\Tr\left(h\gamma\left(\Ket{\Psi}\right)\right),\]
where $h$ is simply the matrix of coefficients $[h_{i,j}]$. The set of all 1RDMs coming from reductions of the pure states (also called the $N$-representable 1RDMs) will be denoted by $\Gamma$, i.e.
\begin{equation}\label{pure-gamma}
\Gamma:=\{\gamma |\ \gamma=\gamma\left(\Ket{\Psi}\right) {\mathrm{\ for\ some\ }}\Ket{\Psi}\}.
\end{equation}
Note that not every $\gamma\in \Gamma$ is the 1RDM of the ground state of $\hat H(h)$ for some $h$. Thus, it is also necessary to consider the set of so-called $v$-representable 1RDMs associated with $\hat W$ which is the set $\Gamma_v\subset \Gamma$ of 1RDMs stemming from ground states of hamiltonians $\hat H(h)$.

Following Levy \cite{levy}, Gilbert \cite{Gilbert} and Lieb \cite{lieb} one defines the RDMFT functional $\mc{F}(\gamma)$ defined on $\Gamma_v$ as follows. Consider the ground state energy of $\hat H(h)$ as a function of $h$, $E_{gs}(h)$. For any $\gamma\in\Gamma_v$ we define
\[\mc{F}(\gamma):=E_{gs}(h)-\Tr\left(h\gamma\right).\]
Then, by the variational principle we have
\begin{equation}\label{search}
E_{gs}(h)=\min_{\gamma\in\Gamma_v}\left(\Tr\left(h\gamma\right)+\mc{F}(\gamma)\right).
\end{equation}
The following two notable problems arise: i) we do not know what the set $\Gamma_v$ is and ii) we do not know what $\mc{F}(\gamma)$ is. Levy has proposed to circumvent this problem by extending the domain of the RDMFT functional and the domain of the search (\ref{search}) to all (possibly non-physical) 1RDMs from $\Gamma$ \cite{levy} and define the extended functional $\tilde{\mc{F}}$ as
\begin{equation}\label{levy-def}
\tilde{\mc{F}}(\gamma):=\min_{\Ket{\Psi}\mapsto \gamma}\Bra{\Psi}\hat W \Ket{\Psi},
\end{equation}
where the minimization is done over all pure states whose 1RDM is $\gamma$. Functional $\tilde{\mc{F}}$ has the following two properties: P1) $\tilde{\mc{F}}(\gamma)+\Tr\left(h\gamma\right)\geq E_{gs}(h)$ for every $h$ and $\gamma\in\Gamma$, P2) $\tilde{\mc{F}}(\gamma_{gs})+\Tr\left(h\gamma_{gs}\right)= E_{gs}(h)$, if $\gamma_{gs}$ is the 1RDM of the ground state of $\hat H(h)$. Properties P1 and P2 imply that for $\gamma\in\Gamma_v$ we have $\tilde{\mc{F}}(\gamma)=\mc{F}(\gamma)$. Thus, functionals $\mc{F}$ as well as $\tilde{\mc{F}}$ are universal, i.e. they are independent of $h$. The do however depend on the interaction $\hat W$. 

Finally, let us point out that by \Eref{levy-def} the functional $\tilde{\mc{F}}(\gamma)$ is upper bounded by $\Bra{\Psi}\hat W \Ket{\Psi}$ for any $\Ket{\Psi}$ whose 1RDM is $\gamma$. We will strongly rely on this elementary fact in the following parts of the paper.

\section{The repulsive gradient in the Bose-Hubbard dimer}
In this section we introduce the main concepts and build some key intuitions that we will develop further in the remaining parts of the paper. A simple and intuitive explanation of the relevant repulsive gradient in the the RDMFT functional that we are going to focus on in this section comes from the Bose-Hubbard dimer. This is a system of two sites labelled by the numbers $i=1,2$ with the corresponding creation operators $b_1^\dagger,b_2^\dagger$ and the interaction given by $\hat W_{BH}=\hat n_1^2+\hat n_2^2$, where $\hat n_i$ is the particle number operator at the site $i$. To simplify things further, we will assume that the system at hand involves only $N=2$ bosons and consider only real wave functions. Then, the corresponding two-boson states can be written in the position basis as the superposition
\[\Ket{\Psi}=a_1\Ket{2_{(1)}}+a_2\Ket{1_{(1)},1_{(2)}}+a_3 \Ket{2_{(2)}},\quad a_1^2+a_2^2+a_3^2=1,\]
where the basis states are $\Ket{2_{(i)}}=\left(b_i^\dagger\right)^2\Ket{{\mathrm{vac}}}$, $\Ket{1_{(1)},1_{(2)}}=b_2^\dagger b_1^\dagger\Ket{{\mathrm{vac}}}$. According to \Eref{1rdm-def} the 1RDM of $\Ket{\Psi}$ is a $2\times 2$ real symmetric matrix whose coefficients are given by
\begin{equation}\label{gamma-dimer}
\gamma\left(\Ket{\Psi}\right)=
\pmatrix{
\gamma_{1,1} & \gamma_{1,2} \cr
\gamma_{1,2} & \gamma_{2,2}
}
=
\pmatrix{
2a_1^2+a_2^2 & \sqrt{2}a_2(a_1+a_3) \cr
\sqrt{2}a_2(a_1+a_3) & a_2^2+2 a_3^2
}.
\end{equation}
Remarkably, the above relations can be inverted in order to express the wave function's coefficients in terms of the variables $\gamma_{i,j}$ leading to an exact expression for the Levy's RDMFT functional according to \Eref{levy-def}. This fact has been studied in the literature -- see \cite{schilling-bosons} for the complete derivation and \cite{cohen1} for a similar treatment of the two-electron Fermi-Hubbard dimer. The resulting expression for the exact RDMFT functional reads
\begin{equation}\label{dimer-exact}
\tilde{\mc{F}}_{BH}(\gamma)=4-\frac{1+\sqrt{1-\gamma_{1,2}^2-(\gamma_{1,1}-1)^2}}{\gamma_{1,2}^2+(\gamma_{1,1}-1)^2},\quad \gamma_{1,2}^2+(\gamma_{1,1}-1)^2\leq 1.
\end{equation}
Obtaining exact expressions for the RDMFT functionals is generally not possible for larger systems as the multi-particle wave functions are parametrised by many more variables than the respective 1RDMs which makes realising the minimization in \Eref{levy-def} in an exact way very difficult.

\begin{figure}[ht]
\centering
\includegraphics[width=\textwidth]{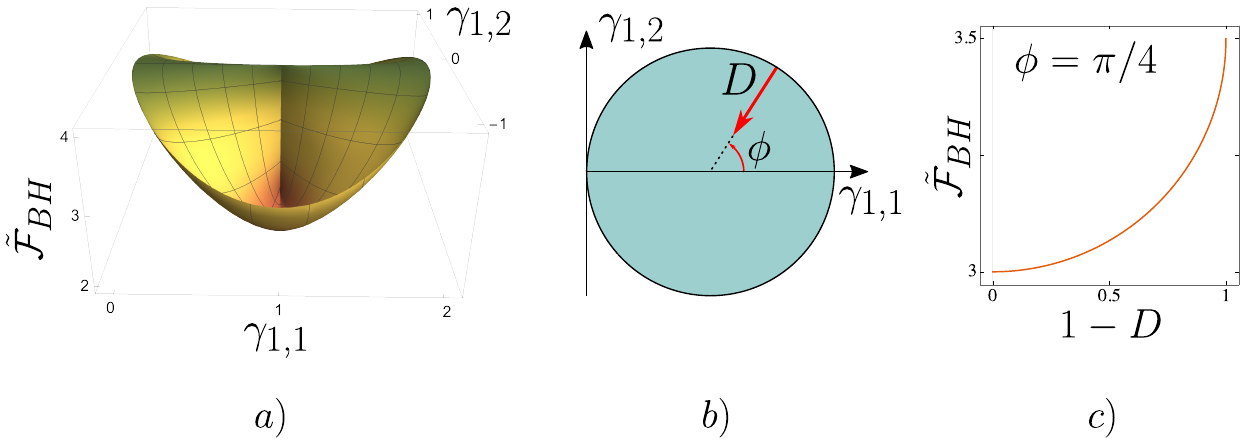}
\caption{The exact RDMFT functional for the $N=2$ Bose-Hubbard dimer. a) The plot of $\tilde{\mc{F}}_{BH}(\gamma)$. b) The domain of the functional is a disk of unit radius which we parametrise by the variables $D$ and $\phi$. c) A radial cut of the plot for $\phi=\pi/4$ showing the cusp singularity of the functional at the boundary of its domain ($D\to 0$).}
\label{dimer-plots}
\end{figure}

According to \Eref{dimer-exact} the domain of the RDMFT functional for the $N=2$ dimer is the disk of unit radius $\gamma_{1,2}^2+(\gamma_{1,1}-1)^2\leq 1$. This fact comes solely from the condition that the 1RDM $\gamma$ has to be positive-semidefinite. Importantly, the 1RDMs which lie on the boundary of the disk have the property that $\gamma^2=2\gamma$, i.e. they have the eigenvalues $\nu_1=2$, $\nu_2=0$. In other words, such 1RDMs come from the states which are completely condensed in a single mode $\ket{\psi}$ being some linear combination of the spatial modes $\ket{1}$ and $\ket{2}$. In this paper, we will be mainly interested in the behaviour of the functional $\tilde{\mc{F}}_{BH}(\gamma)$ when $\gamma$ is close to the set of 1RDMs which come from such completely condensed states. In contrast to the dimer, in higher dimensions the 1RDMs of the completely condensed states form only a subset of the boundary of the domain of the RDMFT functional. However, in the simple example at hand we can visualise the exact functional (Fig.~\ref{dimer-plots}a) and see what exactly is happening in the vicinity of its boundary (Fig.~\ref{dimer-plots}c). Namely, we see that the gradient of $\tilde{\mc{F}}_{BH}(\gamma)$ diverges at the boundary of its domain (this is often called a cusp singularity).

More precisely, we introduce the distance and angle variables, $D\in [0,1]$ and $\phi\in [0,2\pi[$ (Fig.~\ref{dimer-plots}b), which parametrise the disk such that 
\[\gamma_{1,1}=1+(1-D)\cos{\phi},\quad \gamma_{1,2}=(1-D)\sin{\phi},\]
see Fig.~\ref{dimer-plots}a. Using the above parametrisation in \Eref{dimer-exact} we get that 
\[\frac{\partial \tilde{\mc{F}}_{BH}(\gamma(D,\phi))}{\partial D}=-\frac{1-D}{\sqrt{D(2-D)}}\sin^2{\phi},\]
which diverges as $-\frac{\sin^2{\phi}}{\sqrt{2}}\frac{1}{\sqrt{D}}$ in the leading order when $D\to 0$. This repulsive divergence of the RDMFT functional at the boundary of its domain has been confirmed to exist in many more physical systems including the $N$-boson Bose-Hubbard dimer,  homogeneous Bose-Einstein condensates (where it is called the {\it Bose-Einstein condensation force} \cite{schilling-bosons,schilling-bec,carlos-functional}) and translationally invariant one-dimensional fermionic systems \cite{schilling-force} (where it is called the {\it exchange force}). It has been conjectured that it is a property of generic interacting many-body systems. In this paper, we show that this is indeed the case. However, as we are not able to write down the exact functional for a general system and simply compute its derivative, we need to find an appropriate bound for the RDMFT functional whose diverging gradient will imply that the gradient of the exact RDMFT functional diverges as well when $D\to 0$. To this end, for any 1RDM $\gamma$ we construct an ansatz state $\Ket{\Psi(\gamma)}$ such that $\gamma$ is the 1RDM of the state $\Ket{\Psi(\gamma)}$. With such an ansatz at hand, we simply bound the exact functional by
\[\tilde{\mc{F}}(\gamma)\leq \Bra{\Psi(\gamma)}\hat W \Ket{\Psi(\gamma)}\]
and show that the bound diverges repulsively when the (appropriately defined) distance variable $D$ tends to $0$. Let us also remark that the distance variable $D$ has the interpretation as the depletion parameter. To see this, we find the eigenvalues of $\gamma$ to be $\nu_1=2-D$ and $\nu_2=D$. Thus, $D$ is the number of bosons that are outside of the highest occupied mode.

Let us next show how this strategy works out for the $N$-boson Bose-Hubbard dimer. We assume that the 1RDMs are normalised to $\tr\gamma=N$. This implies that the set of 1RDMs is a disk of the radius $N/2$ given by the inequality $(\gamma_{1,1}-N/2)^2+\gamma_{1,2}^2\leq N^2/4$. Thus, we use the parametrisation of $\gamma(D,\phi)$ which reads
\begin{equation}\label{parametrisation}
\gamma_{1,1}=\frac{N}{2}+\left(\frac{N}{2}-D\right)\cos{\phi},\quad \gamma_{1,2}=\left(\frac{N}{2}-D\right)\sin{\phi}.
\end{equation}
We will construct the ansatz state $\Ket{\Psi(\gamma(D,\phi))}\equiv\Ket{\Psi(D,\phi)}$ in the basis of eigenmodes of $\gamma(D,\phi)=(N-D)\Ket{\phi}\!\Bra{\phi}+D\ket{\phi^\perp}\!\bra{\phi^\perp}$, where
\[\Ket{\phi}=\cos{\left(\frac{\phi}{2}\right)}\ket{1}+\sin{\left(\frac{\phi}{2}\right)}\ket{2},\quad \ket{\phi^\perp}=-\sin{\left(\frac{\phi}{2}\right)}\ket{1}+\cos{\left(\frac{\phi}{2}\right)}\ket{2}.\]
The ansatz is the following superposition of the completely condensed state in the mode $\ket{\phi}$ with the doubly-excited state
\[\Ket{\Psi(D,\phi)}=\sqrt{\frac{2-D}{2}}\Ket{N_{(\phi)}}-\sqrt{\frac{D}{2}}\Ket{(N-2)_{(\phi)},2_{(\phi^\perp)}}.\]
It is straightforward to verify that the state $\Ket{\Psi(D,\phi)}$ gives the correct 1RDM (\ref{parametrisation}) which is diagonal in the $\ket{\phi},\ket{\phi^\perp}$ basis and whose diagonal entries in this basis are $N-D$ and $D$. By evaluating the expression $\Bra{\Psi(D,\phi)}\hat W \Ket{\Psi(D,\phi)}$ we get the following upper bound for the RDMFT functional for any $D\leq 2$
\begin{eqnarray*}
\fl \tilde \mc{F}_{BH}(D,\phi)\leq\frac{2-D}{2}\Bra{N_{(\phi)}}\hat W_{BH}\Ket{N_{(\phi)}}-\sqrt{D(2-D)}\Bra{(N-2)_{(\phi)},2_{(\phi^\perp)}}\hat W_{BH}\Ket{N_{(\phi)}} + \\ + \frac{D}{2}\Bra{(N-2)_{(\phi)},2_{(\phi^\perp)}}\hat W_{BH}\Ket{(N-2)_{(\phi)},2_{(\phi^\perp)}}.
\end{eqnarray*}

\begin{figure}[ht]
\centering
\includegraphics[width=.9\textwidth]{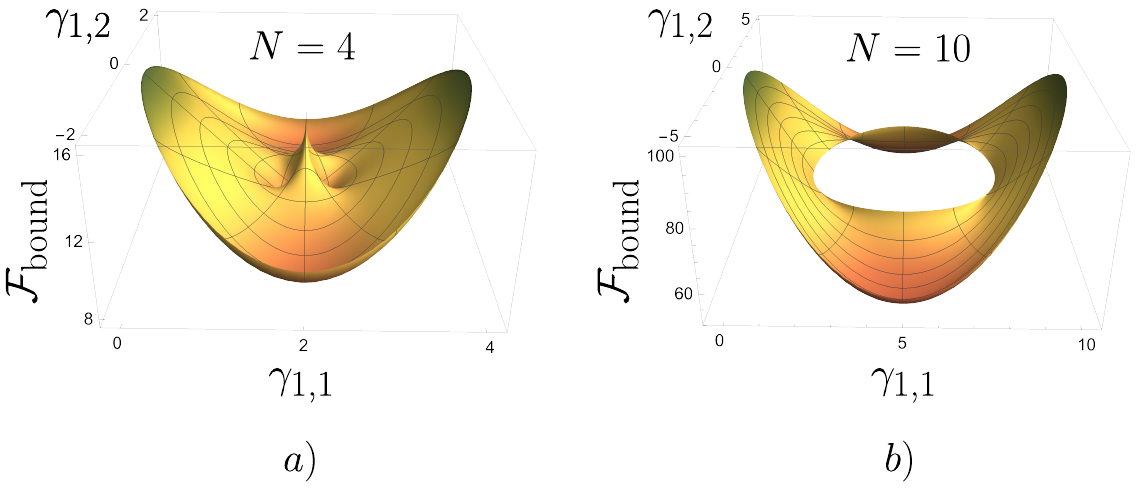}
\caption{Plots of the upper bound (\ref{bound-final}) for the RDMFT functional for the Bose-Hubbard dimer with a) $N=4$ bosons and b) $N=10$ bosons. Note, that the bound is defined only when the distance from the boundary is less than $2$. In particular, this means that whenever $N>4$ the bound is defined on the subset of the set of all the 1RDMs given by the annulus of the inner radius $N/2-2$.}
\label{dimer10plot}
\end{figure}

The upper bound is equal to the exact functional at $D=0$ and its gradient diverges repulsively when $D\to 0$ proportionally to $-1/\sqrt{D}$ in the leading order which implies the desired result. It is however crucial to be assured that $\Bra{(N-2)_{(\phi)},2_{(\phi^\perp)}}\hat W\Ket{N_{(\phi)}}$ is generically strictly positive. Otherwise, the entire argument breaks down as there are no other terms in the bound that give the repulsive gradient. This is done by a straightforward but rather tedious calculation which relies on expressing the interaction $\hat W_{BH}=\hat n_1^2+\hat n_2^2$ in terms of the creation and annihilation operators $b_\phi=\cos(\phi/2)b_1+\sin(\phi/2)b_2$ and $b_{\phi^\perp}=-\sin(\phi/2)b_1+\cos(\phi/2)b_2$. This way, we find that 
\[\Bra{(N-2)_{(\phi)},2_{(\phi^\perp)}}\hat W\Ket{N_{(\phi)}}=\sqrt{\frac{N(N-1)}{2}}\sin^2\phi\geq 0,\]
thus the above overlap vanishes only when $\phi\in\{0,\pi\}$.
A similar calculation allows us to obtain expressions for the remaining expectation values of $\hat W_{BH}$. The resulting upper bound for the RDMFT functional is plotted for $N=4$ and $N=10$ in Fig.~\ref{dimer10plot}. It is given by the formula
\begin{eqnarray}\label{bound-final}
\fl \tilde \mc{F}_{BH}(D,\phi)\leq\frac{1}{4}N\left(1+3N+(N-1)\cos(2\phi)\right)-\sin^2\phi\sqrt{\frac{N(N-1)}{2}}\sqrt{D(2-D)}+ \\ \nonumber
-\frac{D}{2}(N-2)(1+3\cos(2\phi)).
\end{eqnarray}
Finally, the resulting bound for the gradient of the functional at $D\to 0$ reads
\begin{equation}\label{gradient-dimer}
\fl \frac{\partial \tilde{\mc{F}}_{BH}(\gamma(D,\phi))}{\partial D}\leq-\sin^2\phi\sqrt{\frac{N(N-1)}{2}}\frac{1-D}{\sqrt{D(2-D)}}-\frac{1}{2}(N-2)(1+3\cos(2\phi)).
\end{equation}
The above bound for the gradient can be compared with the results of \cite{schilling-bosons} where the first term of the expansion of the gradient in the powers of $D$ has been obtained. The results agree with \Eref{gradient-dimer} up to some factors of $N$ which come from the different choices of the normalisation of 1RDMs.

To end this section, let us remark that analytic bounds for the exact RDMFT functional are potentially very useful tools in approximately solving complex physical systems. This is because the functional is universal for all the single-particle terms $\hat h$ in \Eref{hamiltonian}. For the Hubbard model the single-particle terms include the external on-site potentials and all the possible hopping terms. Thus, quite remarkably, an approximate functional can be applied to the Hubbard model of any geometry. Moreover, finding the ground state energy boils down to the easy task of finding the minimum of a surface (just like the surfaces on Fig.~\ref{dimer10plot} and Fig.~\ref{dimer-plots}a, but generally of a higher dimension) which is tilted by the single-particle terms of the Hamiltonian. In particular, for the dimer this boils down to the almost trivial task of finding the minimum of a two-dimensional surface -- independently of the number of bosons $N$.

\section{The spectral simplex $\Delta_{N,d}$}
\begin{figure}[ht]
\centering
\includegraphics[width=.3\textwidth]{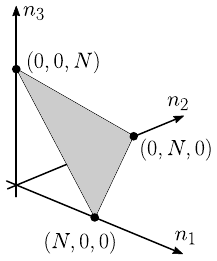}
\caption{The spectral simplex for $d=3$.}
\label{fig:simplex}
\end{figure}

This section is devoted to explaining some key geometric properties of the set $\Gamma$ defined in \Eref{pure-gamma}. One of the main difficulties in RDMFT is the need of optimization over set $\Gamma$ which is in general a complicated non-convex set. However, in contrast to its fermionic counterpart \cite{klyachko,Ruskai70,Ruskai07,BD72}, set $\Gamma$ for bosons is convex and has a tractable description in terms of a geometric object which we call the spectral simplex and denote by $\Delta_{N,d}$. Before we define $\Delta_{N,d}$, we need to take a closer look at a particular set of operations called the single-particle unitaries. The single-particle unitaries are unitary operators generated by single-particle hermitian operators, i.e. they are operations of the form $\hat U=e^{i\hat A}$, where $\hat A=\sum_{i,j=1}^dA_{i,j} b_i b_j^\dagger$ with $A_{i,j}=A_{j,i}^*$ (the star denotes the complex conjugate). The matrix of coefficients $[A_{i,j}]$ will be denoted by $A$. It is a hermitian $d\times d$ matrix, thus $U=e^{iA}$ is a unitary $d\times d$ matrix. Hence, with a given matrix $A$ we have associated a single-particle unitary $\hat U$ and a $d\times d$ unitary $U$. While $\hat U$ acts on $N$-boson wave functions, matrix $U$ acts on 1RDMs via conjugation. Importantly, these two actions are equivalent, namely
\begin{equation}\label{equivariance}
\gamma\left(\hat U\Ket{\Psi}\right)=U\gamma\left(\Ket{\Psi}\right)U^\dagger.
\end{equation}
In particular, any 1RDM $\gamma$ can be diagonalised by the above operations. This means that set $\Gamma$ is parametrised by the spectra of 1RDMs coming from pure states of $N$ bosons. Note that the spectrum of any 1RDM $(n_1,n_2,\dots,n_d)$ has to satisfy
\begin{equation}\label{constraints}
0\leq n_i\leq N\quad {\mathrm{and}}\quad n_1+\dots+n_d=N.
\end{equation} 
These constraints define the spectral simplex $\Delta_{N,d}$ (see Fig~\ref{fig:simplex}).
Simplex $\Delta_{N,d}$ is the convex hull of $d$ vertices $(N,0,\dots,0)$, $(0,N,0,\dots,0)$, $\dots$, $(0,\dots,0,N)$. It turns out that any point of the simplex corresponds to the spectrum of a 1RDM. To see this, consider state 
\[\Ket{\Psi_{\overline \alpha}}:=\sum_{i=1}^d\alpha_i\ket{N_{(i)}},\]
where $\Ket{N_{(i)}}:=\frac{1}{\sqrt{N!}}\left(b_{i}^\dagger\right)^N\Ket{{\mathrm{vac}}}$ and $\overline \alpha=(\alpha_1,\dots,\alpha_d)$ is a vector of real numbers whose squares sum up to one. The 1RDM of $\Ket{\Psi_{\overline \alpha}}$ is diagonal and it reads
\[\gamma\left(\Ket{\Psi_{\overline \alpha}}\right)=\diag\left(N\alpha_1^2,N\alpha_2^2,\dots,N\alpha_d^2\right).\]
Thus, by changing the parameters $\overline\alpha$ we can reach any convex combination of simplex's vertices. Stated another way, the above fact means that any $d\times d$ density matrix $\gamma$ (i.e. a positive-definite hermitian matrix of trace $N$) is the 1RDM of a pure bosonic state of the form $\hat U\Ket{\Psi_{\overline \alpha}}$ for some single-particle $\hat U$ and some $\overline \alpha$. 

For a given typical $d\times d$ density matrix $\gamma$ there exist many $N$-boson pure quantum states that have such a $\gamma$ as their 1RDM. However, this changes diametrically when we consider a special subset of $\Gamma$ which is defined as the set of 1RDMs of completely condensed states of $N$ bosons. All such completely condensed states are of the forms $\hat U\Ket{N_{(1)}}$ where $\hat U$ is a single-particle unitary. By the property (\ref{equivariance}) the 1RDM of a completely condensed state is of the form $U\gamma_{BEC}U^\dagger$, where 
\[\gamma_{BEC}:=\gamma\left(\Ket{N_{(1)}}\right)=\diag\left(N,0,\dots,0\right).\]
Thus, we define
\[\Gamma_{BEC}:=\left\{U\gamma_{BEC}U^\dagger|\ U - d\times d{\mathrm{\ unitary}}\right\}.\]
Importantly, set $\Gamma_{BEC}$ is in a one-to-one correspondence with the set of completely condensed states and the correspondence is $U\gamma_{BEC}U^\dagger\leftrightarrow \hat U\Ket{N_{(1)}}$. Hence, when computing Levy's functional (\ref{levy-def}) no minimization is required and we have 
\begin{equation}
\tilde{\mc{F}}(U\gamma_{BEC}U^\dagger)=\Bra{N_{(1)}}\hat U^\dagger\hat W\hat U \Ket{N_{(1)}}.
\end{equation}

\section{The exposition of the main result}\label{sec:main}
\begin{figure}[ht]
\centering
\includegraphics[width=.5\textwidth]{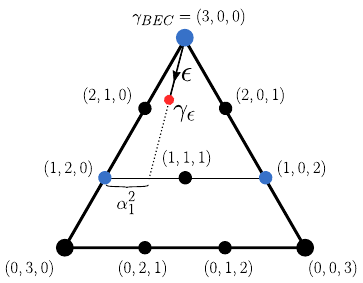}
\caption{The role of parameters $\{\alpha_i\}_{i=1}^{d-1}$ and $\epsilon$ in the ansatz state (\ref{ansatz}) depicted for $d=3$ and $N=3$. The dots denote the spectra of 1RDMs corresponding to the basis states.}
\label{fig:parameters}
\end{figure}

\begin{figure}[ht]
\centering
\includegraphics[width=.4\textwidth]{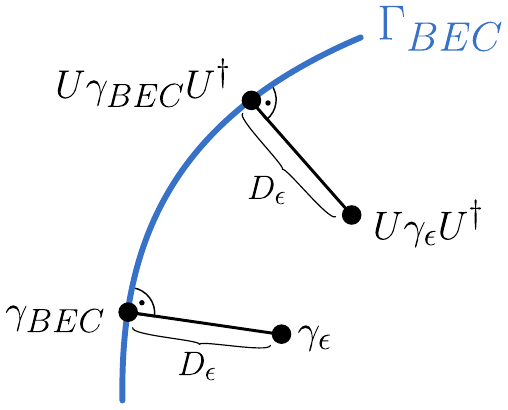}
\caption{The distance $D_\epsilon$ is invariant with respect to conjugation. Matrix $U\gamma_{BEC} U^\dagger$ minimises the distance from $U\gamma_\epsilon U^\dagger $ to the set $\Gamma_{BEC}$. Parameter $\epsilon$ parametrises the shortest path.}
\label{fig:distance}
\end{figure}

The proposed variational ansatz state is a combination of the completely condensed state $\Ket{N_{(1)}}$ and a weighted superposition of states with two bosons outside the condensate. Its precise normalised form reads
\begin{equation}\label{ansatz}
\Ket{\Psi_{\epsilon,\sigma}}=\frac{1}{\sqrt{1+\epsilon^2}}\left(1+\frac{\sigma\epsilon}{\sqrt{2N(N-1)}}b_1^2\sum_{i=1}^{d-1}\alpha_i\left(b_{i+1}^\dagger\right)^2\right)\Ket{N_{(1)}},
\end{equation}
where the variational parameters are $\epsilon\geq0$ and $\sigma=\pm 1$. Numbers $\{\alpha_i\}_{i=1}^{d-1}$ are arbitrary fixed numbers whose role is to probe all directions inside the simplex. They are all nonnegative and satisfy $\alpha_1^2+\dots+\alpha_{d-1}^2=1$. Parameter $\epsilon$ determines the Euclidean distance of $\gamma_\epsilon:=\gamma\left(\Ket{\Psi_\epsilon}\right)$ from the 1RDM of the completely condensed state, $\gamma_{BEC}=\diag\left(N,0,\dots,0\right)$. To see this, recall that the distance between two matrices is given by the Hilbert-Schmidt measure
\begin{equation}\label{HS}
D(\gamma,\gamma')=\sqrt{\Tr\left((\gamma-\gamma')^2\right)}.
\end{equation}
It is straightforward to see that $\gamma_\epsilon$ is diagonal and its precise form reads 
\[\gamma_{\epsilon}=\frac{1}{1+\epsilon^2}\diag\left(N(1+\epsilon^2)-2\epsilon^2,2\epsilon^2\alpha_1^2,\dots,2\epsilon^2\alpha_{d-1}^2\right).\]
Thus, numbers $\{\alpha_i\}_{i=1}^{d-1}$ are in a one-to-one correspondence with the possible directions from vertex $\gamma_{BEC}$ to the interior of the simplex $\Delta_{N,d}$ (see Fig.~\ref{fig:parameters}). 
A straightforward calculation shows that
\begin{equation}\label{dist}
D_{\epsilon}:=D(\gamma_{BEC},\gamma_\epsilon)=\beta_d\frac{\epsilon^2}{1+\epsilon^2},
\end{equation}
where $\beta_d:=2\sqrt{1+\alpha_1^4+\dots+\alpha_{d-1}^4}$. Note that the number of non-condensed bosons in $\gamma_\epsilon$ is exactly $N-N_{BEC}=2\epsilon^2/(1+\epsilon^2)$, thus $D_\epsilon$ has the physical interpretation as the depletion parameter
\[D_{\epsilon}=\frac{1}{2}\beta_d(N-N_{BEC}).\]
Note that distance $D$ is invariant with respect to conjugating its arguments by a $d\times d$ unitary $U$. Hence, $D_{\epsilon}$ is also equal to the distance from $\gamma\left(\hat U\Ket{\Psi_\epsilon}\right)=U\gamma_\epsilon U^\dagger $ to the 1RDM of the corresponding completely condensed state $\gamma\left(\hat U\Ket{N_1}\right)=U \gamma_{BEC} U^\dagger$ (see Fig.~\ref{fig:distance}). 
The ansatz state (\ref{ansatz}) provides an upper bound for the RDMFT functional of the form
\[\tilde{\mc{F}}\left(U\gamma_\epsilon U^\dagger \right)\leq \Bra{\Psi_{\epsilon,\sigma}}\hat U^\dagger \hat W \hat U\Ket{\Psi_{\epsilon,\sigma}}.\]
The above bound is defined for any 1RDM $\gamma$ in the finite region around the set $\Gamma_{BEC}$ consisting of the 1RDMs that satisfy the condition $N-N_{BEC}\leq 2$. This is because any such $\gamma$ can be written as $U\gamma_\epsilon U^\dagger $ for some $U$ and $\gamma_\epsilon$ that are determined via the diagonalisation of $\gamma$. What is more, the matrix $U\gamma_{BEC} U^\dagger$ minimises the distance from $\gamma$ to the set $\Gamma_{BEC}$. The above upper bound has a particularly simple form, namely
\begin{equation}\label{bound-eps}
\Bra{\Psi_{\epsilon,\sigma}}\hat U^\dagger \hat W \hat U\Ket{\Psi_{\epsilon,\sigma}}=\frac{\tilde{\mc{F}}\left(U\gamma_{BEC} U^\dagger \right)+c_1\sigma \epsilon+c_2 \epsilon^2}{1+\epsilon^2}.
\end{equation}
where constants $c_1$, $c_2$ depend on, $N$, $d$, parameters $\{\alpha_i\}_{i=1}^{d-1}$ as well as the precise form of $\hat W$  and the unitary $U$. The expressions for $c_1$ and $c_2$ can be viewed as the following combinations of the two-body integrals $c_1=2\sum_{k=1}^{d-1}\alpha_k\Re I_k$, $c_2=\sum_{k=1}^{d-1}\alpha_k^2I_{k,k}+2\sum_{k<l=1}^{d-1}\alpha_{k,l}\Re I_{k,l}$, where
\begin{eqnarray*}
I_k:=\frac{1}{\sqrt{2N(N-1}}\Bra{N_{(1)}}\hat U^\dagger\hat W\hat U b_1^2\left(b_{k+1}^\dagger\right)^2\Ket{N_{(1)}}, \\
I_{k,l}:=\frac{1}{2N(N-1)}\Bra{N_{(1)}}b_{l+1}^2\left(b_1^\dagger\right)^2\hat U^\dagger\hat W\hat U b_1^2\left(b_{k+1}^\dagger\right)^2\Ket{N_{(1)}}.
\end{eqnarray*}
The two-body integrals can be obtained as closed-form expressions if $d=2$, while for $d>2$ they are computable as convergent series. Importantly, in the following sections we argue that the coefficient $c_1$ is generically nonzero. Using \Eref{dist} this yields the following upper bound for the functional in terms of the distance $D_\epsilon<\beta_d$
\begin{equation}\label{bound}
\fl \tilde{\mc{F}}\left(U\gamma_\epsilon U^\dagger \right)\leq \left(1-\frac{D_\epsilon}{\beta_d}\right)\tilde{\mc{F}}\left(U\gamma_{BEC} U^\dagger \right)+ c_1\sigma \sqrt{\frac{D_\epsilon}{\beta_d}\left(1-\frac{D_\epsilon}{\beta_d}\right)}+c_2 \frac{D_\epsilon}{\beta_d}.
\end{equation}

\begin{figure}[ht]
\centering
\includegraphics[width=.6\textwidth]{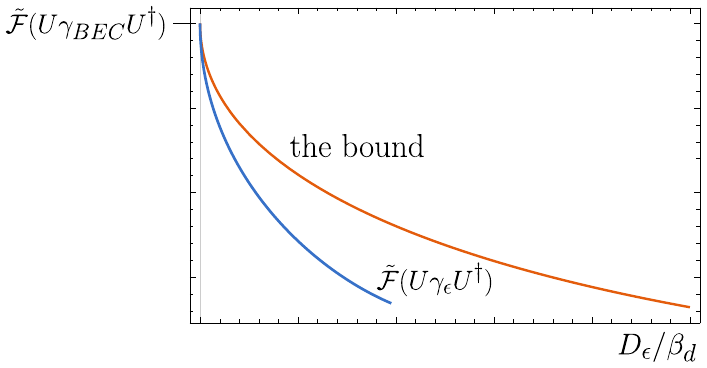}
\caption{The upper bound (\ref{bound}) forces the gradient of the RDMFT functional to diverge repulsively when $D_\epsilon\rightarrow 0^+$.}
\label{fig:bound}
\end{figure}

The final key observation is that the upper bound in \Eref{bound} is exact for $\epsilon=0$. Hence, while approaching the completely condensed state by taking $\epsilon<<1$ (and hence $D_\epsilon<<1$), the gradient of the upper bound with respect to the distance $D_\epsilon$ is also an upper bound for the gradient of the exact functional (see Fig.~\ref{fig:bound}).  Furthermore, the binary parameter $\sigma$ can be chosen so that the gradient of the expression (\ref{bound}) is repulsive for small $D_\epsilon$ and diverges to $-\infty$ as $-\frac{|c_1|}{2\sqrt{\beta_d}}\frac{1}{\sqrt{D_\epsilon}}$ in the leading order, a result anticipated in earlier works \cite{schilling-bosons,schilling-bec}. 

\section{Calculating the coefficient $c_1$}
In this section we provide arguments for the crucial fact that thanks to the specific form of our proposed ansatz state the coefficient $c_1$ in the bound (\ref{bound}) is generically non-zero. 

Although our proof of the bound (\ref{bound}) is elementary and requires only minimal assumptions about the hamiltonian, it becomes heavy with calculations in its most general instance. However, most of the key arguments appear already when considering a dimer, $d=2$. Hence, it will be most instructive to gradually increase the generality of our considerations, beginning with the Bose-Hubbard dimer, then moving to a dimer with an arbitrary pair interaction and finally formulating the proof in its full generality.

\subsection{The Bose-Hubbard dimer}
The interaction term for the Bose-Hubbard hamiltonian on $d$ sites reads 
\[\hat W_{BH}^{(d)}=V\sum_{i=1}^d \hat n_i^2\]
with $V\in {\mathbb R}$ being a real parameter. $V>0$ corresponds to an on-site repulsion, while $V<0$ corresponds to an on-site attraction. The most difficult part of the proof of bound (\ref{bound}) relies on computing $U^\dagger \hat W_{BH} U$. The general strategy is to express the interaction term in terms of generators of the unitary group and then use some well-known facts concerning conjugation of the generators by unitary matrices (in particular, the correspondence with rotations in $(d^2-1)$-dimensions \cite{baker}). Let us next show how this strategy is realized in the case of the Hubbard dimer where $\hat W_{BH}=V(\hat n_1^2+\hat n_2^2)$. Our first goal is to express $\hat W_{BH}$ in terms of the $SU(2)$ generators
\begin{equation}\label{generators2}
\fl \hat X:= b_1^\dagger b_2+b_2^\dagger b_1,\quad \hat Y:= i\left(-b_1^\dagger b_2+b_2^\dagger b_1\right), \quad \hat Z:= b_1^\dagger b_1-b_2^\dagger b_2=\hat n_1-\hat n_2.
\end{equation}
In the following, we will often arrange the above generators into a vector $\hat{\overline R}=\left(\hat R_1,\hat R_2,\hat R_3\right)$, where $\hat R_1=\hat X$, $\hat R_2=\hat Y$, $\hat R_3=\hat Z$. In the above uniform notation, the generators satisfy the standard $SU(2)$ commutation relations, i.e. $[\hat R_i,\hat R_j]=2i\epsilon_{ijk}\hat R_k$, where $\epsilon_{ijk}$ is the Levi-Civita symbol (not to be confused with the real parameter $\epsilon$).

With the above notation established, the Bose-Hubbard interaction term can be written as 
\[\hat W_{BH}^{(2)}=\frac{V}{2}\left(\hat N^2+\hat Z^2\right),\]
where $\hat N$ is the total particle number operator. For $d>2$, we analogously define $\hat Z_{ij}=\hat n_i-\hat n_j$ to obtain
\[\hat W_{BH}^{(d)}=\frac{V}{d}\left(\hat N^2+\sum_{i<j}\hat Z_{ij}^2\right).\]

Let us next take a slight detour and revisit some algebraic properties of conjugation in $d=2$. By definition, any unitary operator for $d=2$ can be written in terms of generators (\ref{generators2}) as $\hat U=e^{i\alpha \overline r\cdotp \hat{\overline R}}$, where $\overline r$ is a vector in three dimensions of unit length. Recall the following key formula for the conjugation in $d=2$
\begin{equation}\label{conjugation2}
e^{-i\alpha \overline r\cdotp \hat{\overline R}}\hat R_je^{i\alpha \overline r\cdotp \hat{\overline R}}=\sum_{i=1}^3C_{i,j}(\alpha,\overline r)\hat R_i.
\end{equation}
Matrix $[C_{i,j}](\alpha,\overline r)$ is real and is well-known to describe a rotation by angle $2\alpha$ about the axis $\overline r$ \cite{baker}. The exact expressions for its entries can be found e.g. in \cite{angular}.

In particular, for the Bose-Hubbard dimer, we have
\[\Bra{\Psi_{\epsilon,\sigma}}\hat U^\dagger \hat W_{BH}^{(2)} \hat U\Ket{\Psi_{\epsilon,\sigma}}=\frac{V}{2}\left(N^2+\Braket{\Phi_{\epsilon,\sigma}^U| \Phi_{\epsilon,\sigma}^U}\right),\]
where $\Ket{\Phi_{\epsilon,\sigma}^U}:=\hat U^\dagger \hat Z \hat U\Ket{\Psi_{\epsilon,\sigma}}$.
Formula (\ref{conjugation2}) applied for $j=3$ gives
\begin{eqnarray*}
\Braket{\Phi_{\epsilon,\sigma}^U | \Phi_{\epsilon,\sigma}^U}=\sum_{i,j=1}^3C_{i,3}C_{j,3}\Bra{\Psi_{\epsilon,\sigma}} \hat R_j\hat R_i\Ket{\Psi_{\epsilon,\sigma}}= \\ =\sum_{i=1}^3C_{i,3}^2\Bra{\Psi_{\epsilon,\sigma}} \hat R_i^2\Ket{\Psi_{\epsilon,\sigma}}.
\end{eqnarray*}
The last line of the above equation follows from the fact that $\Bra{\Psi_{\epsilon,\sigma}} \hat R_j\hat R_i\Ket{\Psi_{\epsilon,\sigma}}=0$ whenever $j=3$ and $i\neq3$ or $i=3$ and $i\neq 3$. This is because $\Ket{\Psi_{\epsilon,\sigma}}$ is a superposition of $\ket{N_1}$ and $b_0^2\left(b_1^\dagger\right)^2\ket{N_1}$, so there have to be two different annihilations in $\hat R_j\hat R_i$ to get a nonzero expectation value. Moreover, $\Ket{\Psi_{\epsilon,\sigma}}$ is a real vector, thus $\Bra{\Psi_{\epsilon,\sigma}} (\hat X\hat Y+\hat Y\hat X)\Ket{\Psi_{\epsilon,\sigma}}=0$. 
Next, we check the following facts by a straightforward calculation
\begin{eqnarray}\label{psieps-action}
\fl \hat X\Ket{\Psi_{\epsilon,\sigma}}=\mc{N}_\epsilon\Big(\left(\sqrt{N}+\sigma\epsilon\sqrt{2(N-1)}\right)\ket{N-1,1}  + \sigma\epsilon\sqrt{3(N-2)}\ket{N-3,3}\Big), \\ \nonumber
\fl\hat Y\Ket{\Psi_{\epsilon,\sigma}}= i\mc{N}_\epsilon\Big(\left(\sqrt{N}-\sigma\epsilon\sqrt{2(N-1)}\right)\ket{(N-1,1} + \sigma\epsilon\sqrt{3(N-2)}\ket{N-3,3}\Big), \\ \nonumber
\fl\hat Z\Ket{\Psi_{\epsilon,\sigma}}=\mc{N}_\epsilon\left(N\ket{N,0}+\sigma\epsilon(N-4)\ket{N-2,2}\right),
\end{eqnarray}
where we used the shorthand notation $\ket{K,L}$ for the normalised state $\left(b_1^\dagger\right)^{K}\left(b_2^\dagger\right)^{L}\Ket{{\mathrm{vac}}}$. The above calculation is key for our proof. Namely, computing expectation values of $\hat R_i^2$ generates terms proportional to $\epsilon$ due to the fact that squares of the form $\left(\sqrt{N}\pm \sigma\epsilon\sqrt{2(N-1)}\right)^2$ appear on the way. Collecting all contributions proportional to $\epsilon$ and $\epsilon^2$, we find the following constants from \Eref{bound-eps}.
\begin{eqnarray*}
\tilde{\mc{F}}\left(U\gamma_{BEC} U^\dagger \right)=\frac{NV}{2}\left(N\left(1+C_{3,3}^2\right)+C_{1,3}^2+C_{2,3}^2\right),\\
c_1=V\sqrt{2N(N-1)}\left(C_{1,3}^2-C_{2,3}^2\right), \\
c_2=\frac{V}{2}\left(N^2+(5N-8)\left(C_{1,3}^2+C_{2,3}^2\right)+(N-4)^2 C_{3,3}^2\right). \\
\end{eqnarray*}
Using the axis-angle representation of the rotation matrix $[C_{i,j}](\overline r,\alpha)$ \cite{angular}, we obtain explicitly
\begin{eqnarray*}
\fl c_1=V\sqrt{2N(N-1)}\big{(}(r_1^2-r_2^2)\left(4 r_3^2 \sin^4(\alpha)-\sin^2(2\alpha)\right)+8 r_1r_2r_3\sin^2(\alpha)\sin(2\alpha)\big{)}.
\end{eqnarray*}

\subsection{The dimer with an arbitrary interaction}
The most general interaction from \Eref{interaction-general} for $d=2$ can always be recast into a combination of $\hat R_i\hat R_j$ with the complex coefficients $\Omega_{i,j}=\omega_{i,j}+i\tau_{i,j}$. Taking into account that the interaction $\hat W$ is hermitian, we have $\hat W=\sum_{i\leq j}\hat W_{i,j}$, where
\[\hat W_{i,j}=\omega_{i,j}\left(\hat R_i\hat R_j+\hat R_j\hat R_i\right)+i\tau_{i,j}\left(\hat R_i\hat R_j-\hat R_j\hat R_i\right).\]
In particular, if $\Omega_{3,3}$ is the only nonzero coefficient, we recover the Bose-Hubbard interaction (up to a constant term).

Applying the conjugation formula (\ref{conjugation2}) to such a most general $\hat W$, we obtain
\begin{equation}\label{uwu-general}
\hat U^\dagger \hat W \hat U=\sum_{k\leq l} \Big(\tilde\omega_{k,l}\left(\hat R_k\hat R_l+\hat R_l\hat R_k\right) + i\tilde\tau_{k,l}\left(\hat R_k\hat R_l-\hat R_l\hat R_k\right)\Big),
\end{equation}
where
\begin{eqnarray*}
\tilde\omega_{k,l}=\sum_{i\leq j}\omega_{i,j}\left(C_{k,i}C_{l,j}+C_{k,j}C_{l,i}\right), \\
\tilde\tau_{k,l}=\sum_{i\leq j}\tau_{i,j}\left(C_{k,i}C_{l,j}-C_{k,j}C_{l,i}\right).
\end{eqnarray*}
The rest of the proof proceeds in a very similar way to the Bose-Hubbard case. Namely, we notice that $\Bra{\Psi_{\epsilon,\sigma}} \left(\hat R_k\hat R_l+\hat R_l\hat R_k\right)\Ket{\Psi_{\epsilon,\sigma}}=0$ whenever $k\neq l$. Similarly, $\Bra{\Psi_{\epsilon,\sigma}} \left(\hat R_k\hat R_3-\hat R_3\hat R_k\right)\Ket{\Psi_{\epsilon,\sigma}}=0$ whenever $k\neq 3$. Moreover, the only terms proportional to $\epsilon$ (i.e. contributing to $c_1$ from \Eref{bound-eps}) come from components of \Eref{uwu-general} with $k=l\neq 3$. Thus, using \Eref{psieps-action} we obtain
\begin{equation*}
c_1=2\sqrt{2N(N-1)}\left(\tilde\omega_{1,1}-\tilde\omega_{2,2}\right).
\end{equation*}
Closed forms of the remaining coefficients from the bound (\ref{bound-eps}) read
\begin{eqnarray*}
\tilde{\mc{F}}\left(U\gamma_{BEC} U^\dagger \right)= N\left(N\tilde\omega_{3,3}+\tilde\omega_{2,2}+\tilde\omega_{1,1}-4\tilde\tau_{1,2}\right)\\
c_2=(5N-8)\left(\tilde\omega_{1,1}+\tilde\omega_{2,2}\right)+\tilde\omega_{3,3}(N-4)^2-2\tilde\tau_{1,2}(N-4).
\end{eqnarray*}

\subsection{The general case}
Our previous considerations concerning an interacting dimer generalise fairly straightforwardly to $d>2$. Firstly, we have more generators of the single-particle unitary operations. Their forms are completely analogous to the dimer case, namely for every pair $1\leq i<j\leq d$ we define the following operators forming the $SU(d)$ algebra
\begin{equation}\label{generatorsd}
\fl \hat X_{i,j}:= b_i^\dagger b_j+b_j^\dagger b_i,\quad \hat Y_{i,j}:= i\left(-b_i^\dagger b_j+b_j^\dagger b_i\right), \quad
\hat Z_{i,j}:= b_i^\dagger b_i-b_j^\dagger b_j=\hat n_i-\hat n_j.
\end{equation}
Thus, we have $3d(d-1)/2$ generators. However, the number of linearly independent generators is $d^2-1$ due to the relations $\hat Z_{i,j}+\hat Z_{j,k}=\hat Z_{i,k}$. We will also arrange all generators in a linear order and use a single greek symbol to enumerate the generators, so that the set of linearly independent generators can be written in a uniform way as $\{\hat R_{\mu}\}_{\mu=1}^{d^2-1}$. Thus, in full analogy to the dimer case, we can write a general pair interaction as $\hat W=\sum_{\mu\leq \nu}\hat W_{\mu,\nu}$, where
\begin{equation}\label{interaction-general}
\hat W_{\mu,\nu}=\omega_{\mu,\nu}\left(\hat R_\mu\hat R_\nu+\hat R_\nu\hat R_\mu\right)+i\tau_{\mu,\nu}\left(\hat R_\mu\hat R_\nu-\hat R_\nu\hat R_\mu\right)
\end{equation}
with $\omega_{\mu,\nu}$ and $\tau_{\mu,\nu}$ being some given real constants.

Conjugating $\hat R_\mu$ by a single-particle unitary of the form $\hat U=e^{i\hat A}$ with $\hat A=\sum_{\mu}a_\mu \hat R_\mu$ results with the transformation
\begin{equation}\label{conjugationd}
e^{-i\hat A}\hat R_\nu e^{i\hat A}=\sum_{\mu=1}^{d^2-1}C_{\mu,\nu}(A)\hat R_\mu,
\end{equation}
where coefficients $C_{\mu,\nu}(A)$ are real and can be computed by taking the exponent of the so-called adjoint action $ad_{A}\left(R_\mu\right):=[A,R_\mu]$. The coefficient matrix $[C_{\mu,\nu}(A)]$ describes a rotation in $d^2-1$ dimensions \cite{baker}. Thus, we get
\begin{equation}\label{uwu-generald}
\hat U^\dagger \hat W \hat U=\sum_{\gamma\leq \delta} \Big(\tilde\omega_{\gamma,\delta}\left(\hat R_\gamma\hat R_\delta+\hat R_\delta\hat R_\gamma\right)+ i\tilde\tau_{\gamma,\delta}\left(\hat R_\gamma\hat R_\delta-\hat R_\delta\hat R_\gamma\right)\Big),
\end{equation}
where
\begin{eqnarray*}
\tilde\omega_{\gamma,\delta}=\sum_{\mu\leq \nu}\omega_{\mu,\nu}\left(C_{\gamma,\mu}C_{\delta,\nu}+C_{\gamma,\nu}C_{\delta,\mu}\right), \\
\tilde\tau_{\gamma,\delta}=\sum_{\mu\leq \nu}\tau_{\mu,\nu}\left(C_{\gamma,\mu}C_{\delta,\nu}-C_{\gamma,\nu}C_{\delta,\mu}\right).
\end{eqnarray*}
Let us remark that the coefficients in \Eref{conjugationd} and \Eref{uwu-generald} depend only on the Lie algebra element $A$, not on the particular representation. In particular, the same equations hold for the fermionic systems described in in Section \ref{sec:singlet}.
 
 The rest of the proof relies on keeping track of contributions proportional to $\epsilon$ and $\epsilon^2$ in expressions of the type $\Bra{\Psi_{\epsilon,\sigma}} \hat R_\gamma\hat R_\delta\Ket{\Psi_{\epsilon,\sigma}}$. The calculation is tedious, but straightforward, thus we move it to \ref{appB}. For the sake of completeness, below we write down the explicit expression for $c_1$ from \Eref{bound-eps}. As it turns out, the only contributions to $c_1$ come from expressions $\Bra{\Psi_{\epsilon,\sigma}} \hat R_\gamma^2\Ket{\Psi_{\epsilon,\sigma}}$, where $\hat R_\gamma=\hat X_{1,k}$ or $\hat R_\gamma=\hat Y_{1,k}$ for some $k$. Thus, the result is completely analogous to the formula describing $c_1$ in the dimer case. Denote by $\omega^{(x)}_k$ the coefficient $\tilde\omega_{\gamma,\gamma}$ in \Eref{uwu-generald} which multiplies $\hat X_{1,k}^2$. Similarly, define $\omega^{(y)}_k$ as the coefficient $\tilde\omega_{\gamma,\gamma}$ in \Eref{uwu-generald} which multiplies $\hat Y_{1,k}^2$. Then, as explained in \ref{appB}
\begin{equation}\label{c1d}
c_1=4\sqrt{2N(N-1)}\sum_{k=1}^{d-1}\alpha_k\left(\omega^{(x)}_k-\omega^{(y)}_k\right),
\end{equation}
where $\{\alpha_k\}_{k=1}^{d-1}$ are parameters of the ansatz state (\ref{ansatz}).

\section{Repulsive gradient in the fermionic RDMFT for the singlet sector}\label{sec:singlet}
\begin{figure}[ht]
\centering
\includegraphics[width=.5\textwidth]{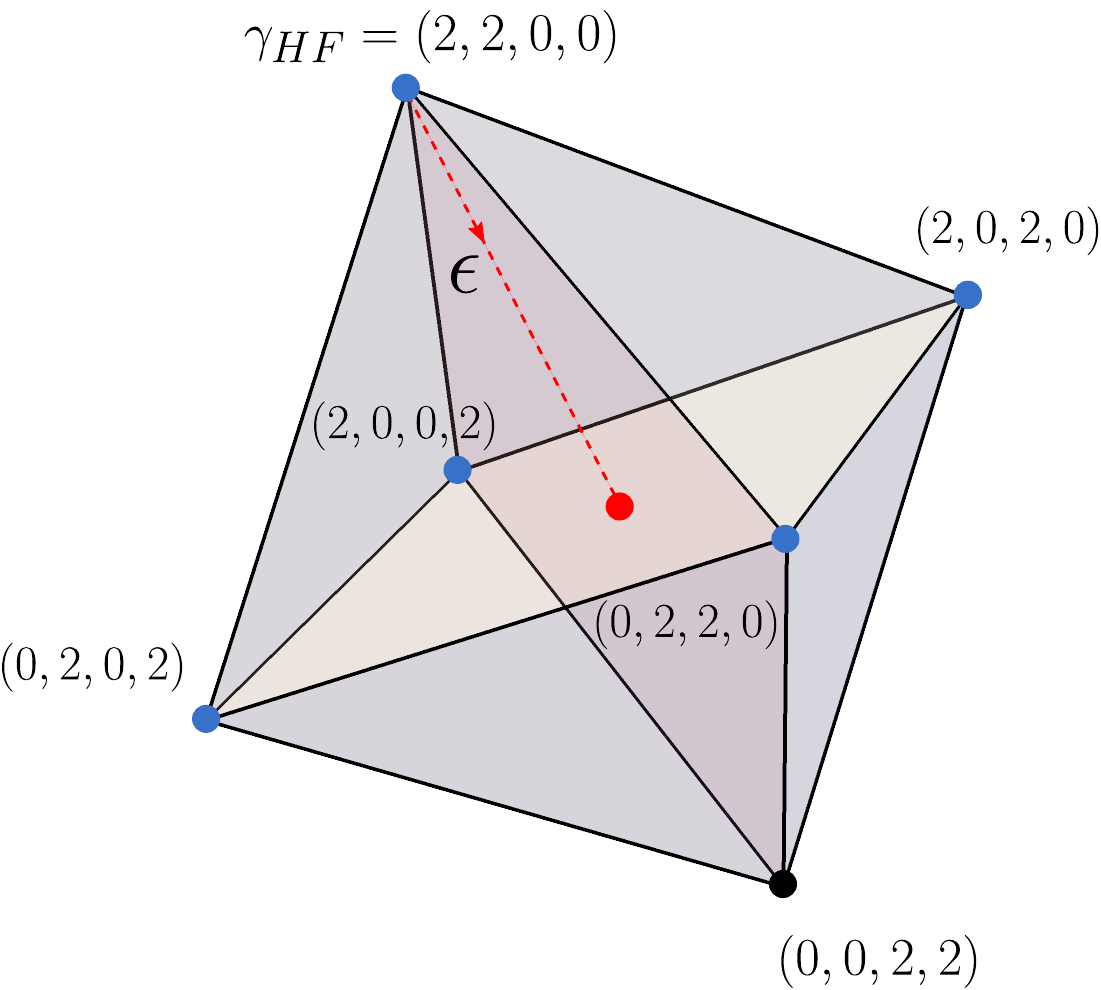}
\caption{The polytope $P_{M,d}$ for $M=2$ and $d=4$. The extreme vertices form an octahedron. The ansatz state (\ref{ansatz-f}) has occupation numbers that lie on the red dashed line connecting the Hartree-Fock point with the red dot defined as the convex combination of the vertices $(2,0,2,0)$, $(0,2,2,0)$, $(0,2,0,2)$ and $(2,0,0,2)$ with the weights $\alpha_{2,3}^2$, $\alpha_{1,3}^2$, $\alpha_{1,4}^2$ and $\alpha_{2,4}^2$ respectively.} 
\label{octahedron}
\end{figure}

In this section we show that an analogous repulsive gradient force is present in systems of interacting spin-$1/2$ fermions for hamiltonians preserving the total spin which we assume to be zero by selecting the singlet sector. The fermions occupy $d$ sites/orbitals with the corresponding annihilation operators $\{a_{i\uparrow},a_{i\downarrow}\}_{i=1}^d$. We start by defining the auxiliary operators $f_{i,j}$ which will be helpful in carrying the calculations in an almost complete analogy to the bosonic case.
\[f_{i,j}:=\sum_{\sigma\in\{\uparrow,\downarrow\}}a_{j\sigma}^\dagger a_{i\sigma} , \quad 1\leq i\leq j\leq d.\]
The one-particle reduced density matrix is given by
\[\gamma\left(\Ket{\Psi}\right)_{i,j}=\sum_\sigma\Bra{\Psi}a_{i\sigma}^\dagger a_{j\sigma}\Ket{\Psi}=\Bra{\Psi}f_{i,j}^\dagger\Ket{\Psi}.\]
The single-particle operators form an $SU(d)$ algebra and are represented on the singlet sector as
\begin{equation}\label{generatorsd-f}
\fl \hat X_{i,j}^f:= f_{i,j}+f_{i,j}^\dagger,\quad \hat Y_{i,j}^f:= i\left(f_{i,j}-f_{i,j}^\dagger\right), \quad
\hat Z_{i,j}^f:= \left[f_{i,j}^\dagger,f_{i,j}\right]=\sum_{\sigma}\left(\hat n_{i\sigma}-\hat n_{j\sigma}\right),
\end{equation}
where $i<j$. As before, for any single-particle unitary $\hat U$ the one-particle reduced density matrices transform equivariantly according to the formula $\gamma\left(\hat U\Ket{\Psi}\right)=U\gamma\left(\Ket{\Psi}\right)U^\dagger$. We also consider the {\it spectral polytope} $P_{M,d}$ which is the polytope formed by the spectra of the one-particle reduced density matrices of the singlet states of $N=2M$ fermions occupying $d$ orbitals (or sites). Importantly, the polytope $P_{M,d}$ has a tractable description \cite{klyachko} -- it is defined only by the Pauli constraints
\[0\leq n_i\leq 2,\quad \sum_{i=1}^dn_i=2M,\]
where $n_i=n_{i\uparrow}+n_{i\downarrow}$ is the number of fermions occupying the $i$th orbital. Clearly, $P_{M,d}$ is not a simplex when $d>3$ or $M>1$. It is a convex hull of $d\choose M$ vertices whose occupation numbers $n_i\in \{0,2\}$ and sum up to $2M$ (see Fig.~\ref{octahedron}).

For $1\leq i_1<i_2<\dots<i_M\leq d$ we define the corresponding singlet Hartree-Fock state as 
\[\Ket{i_1\downarrow,i_1\uparrow,\dots,i_M\downarrow,i_M\uparrow}:=a_{i_1,\downarrow}^\dagger a_{i_1\uparrow}^\dagger\dots a_{i_M\downarrow}^\dagger a_{i_M\uparrow}^\dagger\Ket{\mathrm{vac}}.\]
Any point from the polytope $P_{M,d}$ can be written as the spectrum of the one-particle reduced density matrix of the following combination of the Hartree-Fock states
\[\Ket{\Psi}=\sum_{1\leq i_1<i_2<\dots<i_M\leq d}\alpha_{i_1,\dots,i_M}\Ket{i_1\downarrow,i_1\uparrow,\dots,i_M\downarrow,i_M\uparrow}.\]
The one-particle reduced density matrix corresponding to the above state is easily seen to be diagonal and to correspond to the convex combination of the polytope's vertices with the coefficients $|\alpha_{i_1,\dots,i_M}|^2$.

Any hamiltonian conserving the total spin of $2M$ fermions occupying $d$ orbitals when truncated to the singlet sector can be written as
\begin{equation}\label{ham-f}
\hat H=\sum_{i<j}\left(t_{i,j}\hat X_{i,j}^f+y_{i,j}\hat Y_{i,j}^f+v_{i,j} \hat Z_{i,j}^f\right)+\hat W+\mu\hat N,
\end{equation}
where the interaction term $\hat W$ is an arbitrary linear combination of products of operators (\ref{generatorsd-f}) with real coefficients (i.e. a real polynomial in $\hat X_{i,j}^f$, $\hat Y_{i,j}^f$ and $\hat Z_{i,j}^f$). In particular, the Fermi-Hubbard interaction is obtained as $\hat W_{FH}=\sum_{i<j}V_{i,j}\left(\hat Z_{i,j}^f\right)^2$, $V_{i,j}\in\mathbb{R}$. For the attractive interaction ($V_{i,j}<0$) the Hamiltonian (\ref{ham-f}) with the Bose-Hubbard interaction and only real hopping terms ($y_{i,j}=0$) extended to the entire Fock space is known to have a unique singlet ground state \cite{Lieb-hubbard}. For the repulsive interaction, the ground state is a singlet state only at the half-filling ($N=d$) and only when the graph defined by the hopping geometry is bipartite with the two disjoint sets of vertices having the same cardinality. This is satisfied for instance for the chain-geometry. In what follows, we will assume that $\hat W$ is a pair interaction, i.e. it only consists of terms of degree $2$ in the operators $\hat X_{i,j}^f$, $\hat Y_{i,j}^f$ and $\hat Z_{i,j}^f$.

We consider the following variational ansatz state.
\begin{equation}\label{ansatz-f}
\fl \Ket{\Psi_{\epsilon,\sigma}^f}:=\frac{1}{\sqrt{1+\epsilon^2}}\left(1+\sigma\epsilon\sum_{i=1}^M\sum_{j=M+1}^d\alpha_{i,j}a_{j\downarrow}^\dagger a_{j\uparrow}^\dagger a_{i\uparrow} a_{i\downarrow}\right)\Ket{1\downarrow,1\uparrow,\dots,M\downarrow,M\uparrow},
\end{equation}
where $\sigma\in\{-1,1\}$ and $\alpha_{i,j}$ are real parameters such that $\sum_{i=1}^M\sum_{j=M+1}^d\alpha_{i,j}^2=1$. The one-particle reduced density matrix of the above ansatz state is diagonal and its entries $\gamma\left(\Ket{\Psi_{\epsilon,\sigma}}\right)_{k,k}=\Bra{\Psi_{\epsilon,\sigma}}\hat n_{k\downarrow}+\hat n_{k\uparrow}\Ket{\Psi_{\epsilon,\sigma}}$ read
\begin{eqnarray*}
\gamma\left(\Ket{\Psi_{\epsilon,\sigma}^f}\right)_{k,k}=\frac{2}{1+\epsilon^2}\left(1+\epsilon^2\sum_{j=M+1}^d\alpha_{k,j}^2\right),\quad 1\leq k\leq M, \\
\gamma\left(\Ket{\Psi_{\epsilon,\sigma}^f}\right)_{l,l}=\frac{2\epsilon^2}{1+\epsilon^2}\sum_{i=1}^M\alpha_{i,l}^2,\quad M<l\leq d.
\end{eqnarray*}
Clearly, for $\epsilon=0$ the ansatz state is a Hartree-Fock state and $\gamma\left(\Ket{\Psi_{0,\sigma}^f}\right)=\diag\left(2,\dots,2,0,\dots,0\right)=:\gamma_{HF}$. If $\epsilon>0$, the parameters $\alpha_{i,j}$ swipe through all possible directions coming out of the vertex $\gamma_{HF}$ and going inside the polytope $P_{M,d}$ (see Fig.~\ref{octahedron}). Consequently, for a fixed choice of the coefficients $\alpha_{i,j}$ and for any single-particle unitary $\hat U$ the parameter $\epsilon$ parametrizes the shortest path from the point $U\gamma\left(\Ket{\Psi_{\epsilon,\sigma}^f}\right) U^\dagger$ to the set of the Hartree-Fock states 
\[\Gamma_{HF}:=\left\{U\gamma_{HF}U^\dagger|\ U - d\times d{\mathrm{\ unitary}}\right\}.\]
Alternatively, one may think of the ansatz state $\hat U\Ket{\Psi_{\epsilon,\sigma}^f}$ as having the form (\ref{ansatz-f}) where the annihilation operators $\{a_{i\uparrow},a_{i\downarrow}\}_{i=1}^d$ are replaced by the natural orbital basis annihilators $\{\tilde a_{i\uparrow},\tilde a_{i\downarrow}\}_{i=1}^d$ ordered decreasingly in terms of their corresponding natural occupation numbers. The relevant distance measure is the Hilbert-Schmidt distance from \Eref{HS}. In particular, for any single-particle unitary $\hat U$ we have 
\[D_{\epsilon}:=\min_{\tilde \gamma\in \Gamma_{HF}}D\left(\tilde \gamma,\gamma\left(\hat U\Ket{\Psi_{\epsilon,\sigma}^f}\right)\right)=D\left(\gamma_{HF},\gamma\left(\Ket{\Psi_{\epsilon,\sigma}^f}\right)\right)=\delta_d\frac{\epsilon^2}{1+\epsilon^2},\]
where $\delta_d=2\left(\sum_{i=1}^M\left(\sum_{j=M+1}^d\alpha_{i,j}^2\right)^2+\sum_{j=M+1}^d\left(\sum_{i=1}^M\alpha_{i,j}^2\right)^2\right)^{1/2}$. By evaluating the sum of the $d-M$ lowest natural occupation numbers of the state $\hat U\Ket{\Psi_{\epsilon,\sigma}^f}$, we obtain $N-N_{HF}=\sum_{l=M+1}^d\gamma\left(\Ket{\Psi_{\epsilon,\sigma}}\right)_{l,l}$. By the number $N_{HF}$ we understand the total number of fermions occupying the $M$ most occupied natural orbitals. Using the fact that the squares of the coefficients $\alpha_{i,j}$ sum up to one, we get that $N-N_{HF}=2\epsilon^2/(1+\epsilon^2)$ and thus the distance $D_\epsilon$ has the interpretation as
\[D_\epsilon=\frac{1}{2}\delta_d\left(N-N_{HF}\right).\]
Our goal here is to show that the corresponding fermionic RDMFT functional is upper bounded in the vicinity of the set $\Gamma_{HF}$ by
\begin{equation}\label{bound-f}
\fl \tilde{\mc{F}}\left(U\gamma\left(\Ket{\Psi_{\epsilon,\sigma}^f}\right) U^\dagger \right)\leq \left(1-\frac{D_\epsilon}{\delta_d}\right)\tilde{\mc{F}}\left(U\gamma_{HF} U^\dagger \right)+ c_1^f\sigma \sqrt{\frac{D_\epsilon}{\delta_d}\left(1-\frac{D_\epsilon}{\delta_d}\right)}+c_2^f \frac{D_\epsilon}{\delta_d}.
\end{equation}
By the reasoning presented in Section \ref{sec:main} this implies that the gradient of the RDMFT functional with respect to the distance $D_\epsilon$ diverges repulsively as $-\frac{|c_1^f|}{2\sqrt{\delta_d}}\frac{1}{\sqrt{D_\epsilon}}$ in the vicinity of the Hartree-Fock point $U\gamma_{HF}U^\dagger$. To this end, we show that the following holds
\begin{equation}\label{bound-eps-f}
\Bra{\Psi_{\epsilon,\sigma}^f}\hat U^\dagger \hat W \hat U\Ket{\Psi_{\epsilon,\sigma}^f}=\frac{\tilde{\mc{F}}\left(U\gamma_{HF} U^\dagger \right)+c_1^f\sigma \epsilon+c_2^f \epsilon^2}{1+\epsilon^2}
\end{equation}
for any pair-interaction $\hat W$. As before, the parameters $c_1^f$ and $c_2^f$ are linear combinations of the appropriate two-body integrals. For the most general interaction given by \Eref{interaction-general} the proof of \Eref{bound-eps-f} is done using the conjugation formula (\ref{uwu-generald}). The details are deferred to \ref{appF}. As it turns out, the only contributions to $c_1^f$ in the LHS of \Eref{bound-eps-f} come from the expressions $\Bra{\Psi_{\epsilon,\sigma}} \hat R_\gamma^2\Ket{\Psi_{\epsilon,\sigma}}$, where $\hat R_\gamma=\hat X_{k,l}^f$ or $\hat R_\gamma=\hat Y_{k,l}^f$ for $1\leq k\leq M$ and $l>M$. Thus, the result is completely analogous to the formula describing $c_1$ in the bosonic case. Denote by $\omega^{(x)}_{k,l}$ the coefficient $\tilde\omega_{\gamma,\gamma}$ in \Eref{uwu-generald} which multiplies $\left(\hat X_{k,l}^f\right)^2$. Similarly, define $\omega^{(y)}_{k,l}$ as the coefficient $\tilde\omega_{\gamma,\gamma}$ in \Eref{uwu-generald} which multiplies $\left(\hat Y_{k,l}^f\right)^2$. Then, 
\begin{equation}\label{c1d}
c_1^f=8\sum_{k=1}^{M}\sum_{l=M+1}^d\alpha_{k,l}\left(\omega^{(x)}_{k,l}-\omega^{(y)}_{k,l}\right),
\end{equation}
where $\alpha_{k,l}$ are the parameters of the ansatz state (\ref{ansatz-f}). Thus, the coefficient $c_1^f$ is generically nonzero.

\subsection{The form of the approximate functional}
For any 1RDM $\gamma$ the approximate functional is computed from the ansatz state (\ref{ansatz-f}) by using the one-particle basis of the natural orbitals corresponding to the decreasingly ordered spectrum (i.e. the natural occupation numbers) of $\gamma$. Denote by $2\geq \nu_1\geq \nu_2\geq\dots\geq \nu_d\geq 0$ the decreasingly ordered natural occupation numbers of $\gamma$. Then, $\gamma$ is a $d\times d$ matrix $\gamma=\sum_{i=1}^d \nu_i\Ket{\phi_i}\!\Bra{\phi_i}$ and we define the operators $\{\tilde a_{i\downarrow},\tilde a_{i\uparrow}\}_{i=1}^d$ so that they refer to the annihilation operators corresponding to the natural orbitals $\{\Ket{\phi_i}\}_{i=1}^d$. According to the \Eref{ansatz-f}, after slight modifications, the corresponding ansatz state reads
\begin{equation}\label{ansatzNO}
\fl \Ket{\Psi_{\epsilon}^f(\gamma)}:=\frac{1}{\sqrt{1+\epsilon^2}}\left(1+\epsilon\sum_{i=1}^M\sum_{j=M+1}^d s_{i,j}\sqrt{\alpha_{i,j}}\tilde a_{j\downarrow}^\dagger \tilde a_{j\uparrow}^\dagger \tilde a_{i\uparrow} \tilde a_{i\downarrow}\right)\Ket{HF(\gamma)},
\end{equation}
where $\Ket{HF(\gamma)}:=\Ket{\phi_1\downarrow,\phi_1\uparrow,\dots,\phi_M\downarrow,\phi_M\uparrow}$ and $\alpha_{i,j}\geq 0$, $\sum_{i=1}^M\sum_{j=M+1}^d\alpha_{i,j}=1$ and $s_{i,j}\in\{-1,1\}$. The coefficients $\{\alpha_{i,j}\}$ and $\epsilon$ are determined by demanding the 1RDM of  $\Ket{\Psi_{\epsilon,\sigma}^f(\gamma)}$ to have the natural occupation numbers $\{\nu_i\}_{i=1}^d$. This leads to 
\[\epsilon=\left(\frac{2}{N-N_{HF}}-1\right)^{-1/2},\quad N_{HF}:=\sum_{i=1}^M\nu_i\]
and to the following linear equations for the $\alpha_{i,j}$'s:
\begin{eqnarray}\label{alpha-eqs}
\fl \sum_{j=M+1}^d\alpha_{k,j}=1-\frac{2-\nu_k}{\sum_{i=M+1}^d\nu_i},\quad 1\leq k\leq M, \quad
\sum_{i=1}^M\alpha_{i,l}=\frac{\nu_l}{\sum_{i=M+1}^d\nu_i},\quad M<l\leq d
\end{eqnarray}
on top of the normalization. When $d>3$, the solutions of the equations (\ref{alpha-eqs}) together with the normalization and the non-negativity conditions form a convex set of a positive dimension as the number of unknowns exceeds the number of equations. Let us denote the space of $\{\alpha_{i,j}\}$ satisfying the equations (\ref{alpha-eqs}) and the conditions $\alpha_{i,j}\geq 0$ by $\Delta(\overline\nu)$. The set $\Delta(\overline\nu)$ is a simplex of dimension $(d-M)(M-1)-M$. The approximate functional is computed as the minimum of the expectation value $\Bra{\Psi_{\epsilon}^f(\gamma)}\hat W\Ket{\Psi_{\epsilon}^f(\gamma)}$ over all solutions from the simplex $\Delta(\overline\nu)$. This ultimately boils down to the following expression
\begin{eqnarray}\label{functional-f}
\fl G(\gamma)=\min_{\{\alpha_{i,j}\}\in \Delta(\overline\nu)}\Bigg{\{}\frac{1}{1+\epsilon^2}\Big{(}I_0+2\epsilon\sum_{i=1}^M\sum_{j=M+1}^d s_{i,j}\sqrt{\alpha_{i,j}}\Re I_{(i,j)}+
 \\ \nonumber+\epsilon^2\sum_{i,k=1}^M\sum_{j,l=M+1}^ds_{i,j}s_{k,l}\sqrt{\alpha_{i,j}\alpha_{k,l}}I_{(i,j),(k,l)}\Big{)}\Bigg{\}},
\end{eqnarray}
where the two-fermion integrals read
\begin{eqnarray*}
\fl I_0(\gamma):=\Bra{HF(\gamma)}\hat W\Ket{HF(\gamma)}, \quad I_{(i,j)}(\gamma)=\Bra{HF(\gamma)}\hat W\tilde a_{j\downarrow}^\dagger \tilde a_{j\uparrow}^\dagger \tilde a_{i\uparrow} \tilde a_{i\downarrow}\Ket{HF(\gamma)}, \\
I_{(i,j),(k,l)}(\gamma)=\Bra{HF(\gamma)} \tilde a_{k\downarrow}^\dagger \tilde a_{k\uparrow}^\dagger \tilde a_{l\uparrow} a_{l\downarrow}\hat W\tilde a_{j\downarrow}^\dagger \tilde a_{j\uparrow}^\dagger \tilde a_{i\uparrow} \tilde a_{i\downarrow}\Ket{HF(\gamma)}
\end{eqnarray*}
and 
\[s_{i,j}=-{\mathrm{sgn}}\left( \Re I_{(i,j)}(\gamma)\right).\]
Thus, the computation of $G(\gamma)$ requires performing a convex minimization over the set $\Delta(\overline\nu)$. Importantly, the region of applicability of the approximate functional is given by the condition $N_{HF}\geq N-2$. This is always satisfied when $N\leq 4$, however when $N>4$ the approximate functional is defined only on a subset of the set of all the $N$-representable 1RDMs. Thus, we expect the approximate functional to perform best for low particle numbers.

For the dimer ($d=2$), there is no convex optimization required and the functional (\ref{functional-f}) simplifies to
\begin{equation}\label{functionald2}
G(\gamma)=\frac{1}{1+\epsilon^2}\Big{(}I_0-2\epsilon\left|\Re I_{2,2}\right|+\epsilon^2 I_{(2,2),(2,2)}\Big{)}.
\end{equation}
In fact, this is the exact RDMFT functional whenever $\gamma\neq \diag(1,1)$. This is because any singlet state of two spin-$1/2$ fermions with $d=2$ whose one-body reduced density matrix is not the identity matrix necessarily has the form (\ref{ansatzNO}) up to a relative phase, when written in the natural orbital basis. The expression (\ref{functionald2}) is equivalent to the expressions for the exact density functional studied in other works \cite{cohen1,schilling-bosons,pastor}.  Another simplification arises when $N=2$, regardless of $d$. Then, the coefficients $\{\alpha_{1,j}\}_{j=2}^d$ follow immediately from the \Eref{alpha-eqs} as 
\[\alpha_{1,j}=\frac{\nu_j}{\sum_{i=M+1}^d\nu_i}.\]

\subsection{Electron transfer in the Fermi-Hubbard model}
The repulsive Fermi-Hubbard chain is a good testing ground for our proposed approximate functional. The ground state of $N$ electrons on the Fermi-Hubbard chain of the length $d=2N$ (half-filling) is known to be a singlet state \cite{Lieb-hubbard}. The Hamiltonian reads
\[\fl \hat H_{FH}=\sum_{i=1}^{d-1} t_{i,i+1}\sum_{\sigma\in\{\uparrow,\downarrow\}}\left(a_{i\sigma}^\dagger a_{(i+1)\sigma}+a_{(i+1)\sigma}^\dagger a_{i\sigma}\right)+\sum_{i=1}^d \sum_{\sigma\in\{\uparrow,\downarrow\}} v_i n_{i\sigma}+\sum_{i=1}^d V_{i}n_{i\uparrow}n_{i\downarrow}\]
with $V_i>0$ for all $i$. In the calculations to follow, we will assume an open chain with i) uniform hopping terms, i.e. $t_{i,i+1}\equiv-t$ with $t\geq 0$ for all $i$, ii) uniform interaction strength, i.e. $V_i\equiv V$ for all $i$ and iii) the on-site potentials $v_i$ that change by even increments from $v_1=0$ to $v_d=v$, i.e. $v_i=v(i-1)/(d-1)$. Then, up to a polynomial term in the total number of electrons, the hamiltonian $\hat H_{FH}$ is easily seen to have the form
\begin{equation}\label{fh-ham}
\hat H_{FH}=-t\sum_{i=1}^{d-1}\hat X_{i,i+1}^f+\frac{v}{2}\sum_{i=1}^{d-1}\frac{i(i-d)}{d-1}\hat Z_{i,i+1}^f+\frac{1}{2d}V\sum_{i<j=1}^d\left(\hat Z_{i,j}^f\right)^2.
\end{equation}

The exact RDMFT functional in the dimer case ($d=2$) has been studied in \cite{cohen1} where its performance has been compared with other approximate density functionals. In particular, the authors of \cite{cohen1} studied the electron transfer problem which asks about how the diagonal entries of the 1RDM of the ground state for the Hamiltonian (\ref{fh-ham}) depend on the on-site potential difference $v$ between the endpoints of the chain. It has been shown that the M\"{u}ller and Power approximate density functionals \cite{muller,power} give qualitatively incorrect results for the electron transfer, where the failure is most visible in the strong interaction limit ($t>> V$). Other works show that approximate molecular density functionals generally fail to give the qualitatively correct description of the electron transfer and point out that this is an important challenge to improve the performance of the approximate functionals in this area \cite{transfer,challenges}. In the following part of this section, we show that out proposed approximate density functional does indeed perform well in the electron transfer calculations of small Hubbard chains. In particular, we show that our functional gives accurate values of the ground state energy for the two-electron singlet systems and that show that it describes the electron transfer with a remarkable accuracy for the four-electron chain with four sites.

\begin{figure}[ht]
\centering
\includegraphics[width=\textwidth]{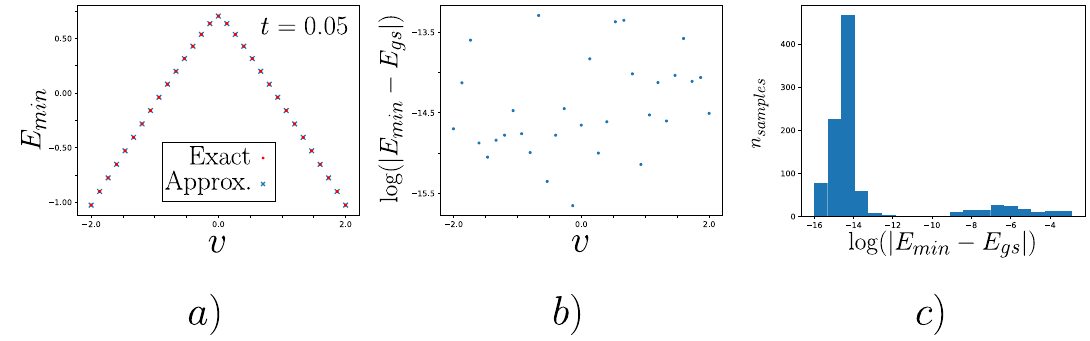}
\caption{The approximate functional (\ref{functional-f}) gives very accurate values for the ground state energy when $N=2$ for the uniform Fermi-Hubbard interaction. a) The energy $E_{min}=G(\gamma_{gs})+\tr h\gamma_{gs}$ (blue crosses) vs. the exact ground state energy $E_{gs}$ (red dots) for the chain of length $d=20$ with the Hamiltonian (\ref{fh-ham}). b) The absolute errors between $E_{min}$ and $E_{gs}$ from the plot a). c) The distribution of the ground state energy errors for the Fermi-Hubbard model with $d=6$ sites with all the hopping amplitudes nonezro. The hopping amplitudes and the on-site potentials were randomly sampled $10^3$ times. The resulting mean error was of the order $10^{-(13\pm 3)}$. The maximum error for the sample was of the order $10^{-3}$, but for $77\%$ of the sample the error was of the order $10^{-14}$ or less.}
\label{N2-conj}
\end{figure}

By applying the functional (\ref{functional-f}) to the two-electron ($N=2$) Fermi-Hubbard chain, we observed numerically that for the uniform interaction and for all the sampled values of the parameters $t\in[0,1]$ and $v\in [-2,2]$ of the Hamiltonian (\ref{fh-ham}), the approximate functional $G$ agreed with the exact functional within the numerical accuracy. In other words, the ground state energy error $\left| G(\gamma_{gs})+\tr h\gamma_{gs}-E_{gs}\right|$ was of the mean order $10^{-14.5}$ (see Fig.~\ref{N2-conj}a and Fig.~\ref{N2-conj}b) which is the same as the error of the $LU$ diagonalization algorithm which was used in the calculation of $E_{gs}$. Thus, the approximate functional coincides with the exact functional for the Hubbard chain with the Hamiltonian (\ref{fh-ham}) within the numerical accuracy range. We also tested other geometries of the Hubbard model. In particular, for the  Fermi-Hubbard model on $d=6$ sites with a uniform repulsive interaction we sampled $10^3$ values of the hopping and the on-site potentials. There, the hopping and the on-site potentials were allowed to be arbitrary and they did not correspond to the chain geometry. The above parameters were sampled from the flat distributions $t_{i,j}\in[-1,1]$ and $v_i\in [-2,2]$. For each set of the sampled parameters a minimization of the approximate density functional $G(\gamma)$ has been performed. This showed the mean energy error of the order $10^{-(13\pm 3)}$ (see Fig.~\ref{N2-conj}c). This number is to be compared with the error of the $LU$ diagonalization algorithm which was used in the calculation of $E_{gs}$ which was of the order $10^{-15}$ or with the typical difference between the lowest and the highest eigenvalue of the random Hamiltonian which was of the order $10^1$. 

The above good results for the ground state energy of the two-electron systems are due to the fact that the excitations of the tested systems exhibit localization in the natural orbital basis making the superposition of doubly-excited Slater determinants in the ansatz state (\ref{ansatz-f}) an accurate guess for the ground state of the system. The ground states of systems of $N>2$ electrons  generally involve higher excitations and thus the correlation energy is expected to be recovered by our proposed functional only to a small extent. This can be seen already in the half-filled chain of length four ($N=d=4$) studied in the remaining part of this section. However, as the numerical results show, the doubly-excited ansatz is enough to recover the electron transfer in the ground state of this system.

\begin{figure}[ht]
\centering
\includegraphics[width=\textwidth]{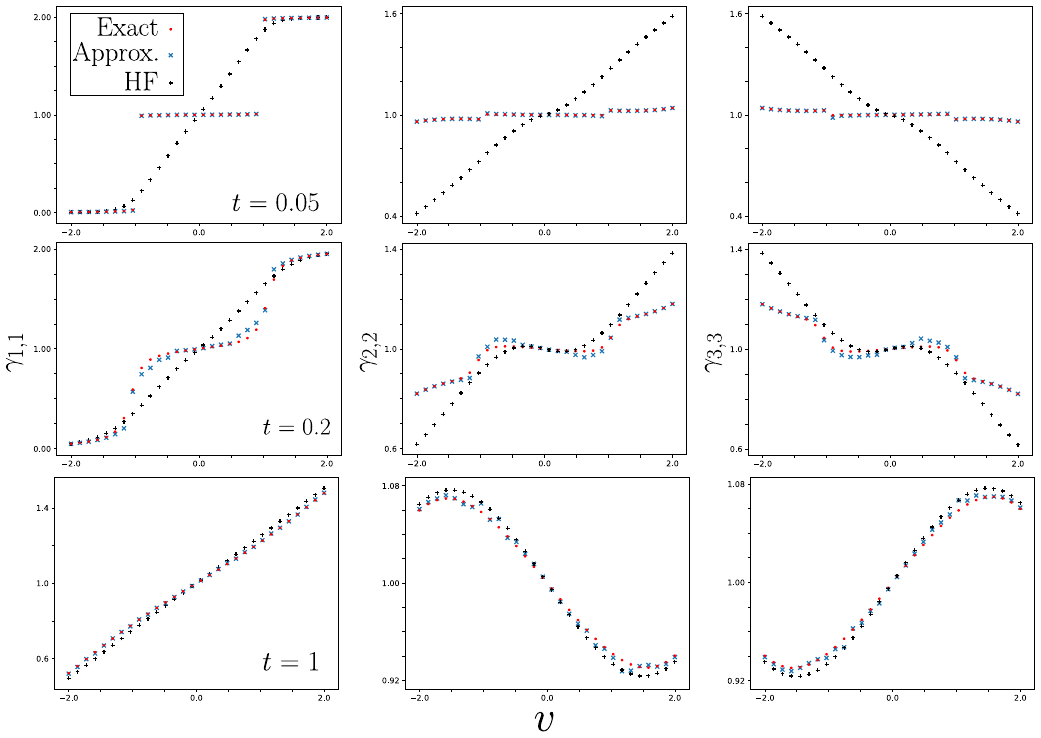}
\caption{The electron transfer for the Fermi-Hubbard chain with $d=4$ sites and $N=4$ electrons with the Hamiltonian (\ref{fh-ham}) computed using the exact diagonalization (red dots), the Hartree-Fock method (black pluses) and the approximate functional (blue crosses). Dependence of $\gamma_{i,i}$ on the on-site potential difference $v$ is plotted for the chosen values of the hopping amplitude: $t=0.05$ (strong interaction), $t=0.2$ (medium interaction) and $t=1$ (weak interaction). The top and the middle plots indicate the failure of the Hartree-Fock method and the good accuracy of the approximate functional.}
\label{transfer4}
\end{figure}

\begin{figure}[ht]
\centering
\includegraphics[width=\textwidth]{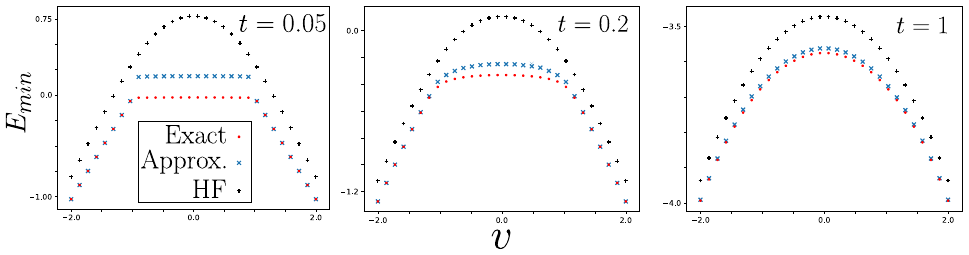}
\caption{The minimised energy vs. the on-site potential difference $v$ for the half-filled Fermi-Hubbard chain with $d=4$ sites with the Hamiltonian (\ref{fh-ham}) computed using the exact diagonalization (red dots), the Hartree-Fock method (black pluses) and the approximate functional (blue crosses). The approximate functional generally does not recover the correlation energy. This is most visible in the strongly correlated limit ($t=0.05$) for the values of $v$ close to $0$. There, the fraction of the recovered correlation energy is only about $70\%$. }
\label{energy4}
\end{figure}

For $N=d=4$ \Eref{alpha-eqs} leads to the following expressions for the coefficients $\alpha_{1,3},\alpha_{1,4},\alpha_{2,3},\alpha_{2,4}$ of the ansatz state (\ref{ansatzNO}).
\begin{eqnarray*}
\frac{2-\nu_2-\nu_4}{\nu_3+\nu_4}\leq \alpha_{1,3}\leq \min\left\{\frac{\nu_3}{\nu_3+\nu_4},\frac{2-\nu_2}{\nu_3+\nu_4}\right\},\quad 
\alpha_{1,4}=\frac{2-\nu_2}{\nu_3+\nu_4}-\alpha_{1,3}, \\
\alpha_{2,3}=\frac{\nu_3}{\nu_3+\nu_4}-\alpha_{1,3},\quad \alpha_{2,4}=\alpha_{1,3}-\frac{2-\nu_2-\nu_4}{\nu_3+\nu_4}.
\end{eqnarray*}
Thus, the minimization in \Eref{functional-f} is done simply over the interval containing the coefficient $\alpha_{1,3}$. The results for the ground state energy and the electron transfer are plotted in Fig.~\ref{transfer4} and Fig.~\ref{energy4}, where the interaction strength has been fixed to $V=1$ due to the fact that the physically meaningful quantity is the ratio $t/V$. As we mentioned before, our approximate functional gives a qualitatively correct electron transfer with a remarkably good accuracy, especially in the limit of the strong interaction. On the other hand, the correlation energy is recovered only to a small extent as can be seen in Fig.~\ref{energy4}.

{\it Technical note: the numerical minimization of $G(\gamma)$ over the set of 1RDMs has been done using the adaptive Nelder-Mead method \cite{NM1,NM2} where the convergence treshold was the difference in function's arguments between iterations which was set to $10^{-8}$. Typically, the convergence was attained after about $10^4$ iterations.}

\section{Summary and discussion}\label{sec:discussion}
Using the ansatz states (\ref{ansatz}) and (\ref{ansatz-f}) we have constructed universal upper bounds (\ref{bound}) and (\ref{bound-f}) for the bosonic and fermionic RDMFT functionals in a finite region around of the set of fully condensed (respectively the Hartree-Fock) states for any pair interaction. The upper bounds depend on the (generalised) depletion parameter which is the number of non-condensed bosons, $N-N_{BEC}$, or the number of fermions which do not occupy the $N/2$ highest occupied natural orbitals, $N-N_{HF}$. Because the upper bound coincides with the exact RDMFT functional when $N_{BEC}=N$ (all the bosons condensed) or $N_{HF}=N$ (system in a Hartree-Fock state), the gradient of the upper bound is also an upper bound for the gradient of the exact functional with respect to the depletion parameter. The gradient is shown to diverge repulsively when $N_{BEC}\to N$ ($N_{HF}\to N$ respectively). This provides an alternative explanation for the existence of the quantum depletion effect which uses only the very fundamental properties of the quantum theory: the variational principle and the geometry of the set of one-particle reduced density matrices. 

We also tested our proposed approximate functional by computing the electron transfer in small Hubbard chains involving $N=2$ and $N=4$ electrons. We showed that in these systems the functional performs remarkably better that some other widely used approximate functionals, and thus our results shed some light on the possible ways of improving the performance of the density matrix functional  methods in the electron transfer computations. In particular, the ansatz states (\ref{ansatz}) and (\ref{ansatz-f}) take into account only double excitations in the natural orbitals. Such ansatz states are expected to work well only in small systems (like the ones tested in this paper) or in extremely inhomogeneus ones where the excitations are effectively localised in a small part of the natural orbital basis. Our functional seems to perform well only in the specific task of computing the electron transfer, as it generally does not recover the correlation energy in the strong interaction limit, as seen in Fig.~\ref{energy4}. A possible way of constructing more accurate functionals would be to superpose Slater determinants with higher numbers of excitations, as one expects that the number of excitations grows linearly with the size of the system. Finally, another practical potential use for our results could be a testing criterion for finding accurate density functionals -- an accurate approximate density functional should reproduce the repulsive gradient. 

Let us remark that one of the main obstructions in finding the exact RDMFT functional is that it requires a detailed knowledge of the geometry of the set of single-particle reduced density matrices. The description of this set has been long-recognised as one of the most important fundamental problems in quantum chemistry \cite{coleman}. The solution of this problem in the late 2000s \cite{klyachko} provided new insights to the RDMFT and it may lead to describing the counterparts of the quantum depletion effect for more general systems of fermions or distinguishable particles using the generalised doubly-excited states \cite{doubly}. Finally, note that throughout the paper we are implicitly assuming that the Levy's RDMFT functional $\tilde{\mc{F}}$ is actually differentiable at points from the set $\Gamma_{BEC}$ ($\Gamma_{HF}$). Earlier works studying the exact RDMFT functional for Hubbard dimers \cite{schilling-bosons,cohen1} suggest that this is a reasonable assumption. However, we are not aware of any mathematically rigorous proof of the fact that the extended functional is differentiable at $\Gamma_{BEC}$ ($\Gamma_{HF}$) for any pair interaction $\hat W$. Perhaps so-called symplectic slice techniques used in \cite{doubly} may shed some light on a solution of this problem.

\section*{Acknowledgments}
I would like to thank Christian Schilling for introducing me to the subject of the Reduced Density Matrix Functional Theory and for the feedback at the early stages of the manuscript. I am also very grateful to Jonathan Robbins for providing careful feedback about the manuscript and for many useful discussions.

\appendix
\section{Calculation of $c_1$ for $d\geq 2$}\label{c1-app}

\subsection{The bosonic density functional}\label{appB}

Recall the definition of our ansatz state
\begin{equation}\label{ansatz-app}
\Ket{\Psi_{\epsilon,\delta}}=\mc{N}_\epsilon\left(\Ket{N_{(1)}}+\sigma\epsilon\sum_{i=1}^{d-1}\alpha_i\Ket{(N-2)_{(1)},2_{(i+1)}}\right),
\end{equation}
where $\epsilon\geq 0$, $\alpha_1^2+\dots+\alpha_{d-1}^2=1$, $\sigma=\pm 1$ and the normalization constant reads $\mc{N}_\epsilon=1/\sqrt{1+\epsilon^2}$. The basis states are defined in terms of bosonic creation and annihilation operators
\[[b_i,b_j^\dagger]=\delta_{i,j}, \quad [b_i^\dagger,b_j^\dagger]=[b_i,b_j]=0, \quad \hat n_i:= b_i^\dagger b_i\]
as 
\[\fl\Ket{(N_1)_{(1)},(N_2)_{(2)},\dots,(N_d)_{(d)}}:=\frac{1}{\sqrt{N_1!N_2!\dots N_d!}}\left(b_1^\dagger\right)^{N_1}\left(b_2^\dagger\right)^{N_2}\dots\left(b_d^\dagger\right)^{N_d}\Ket{{\mathrm{vac}}}.\]
We also have
\begin{eqnarray*}
\fl b_i\Ket{(N_1)_{(1)},\dots,(N_i)_{(i)},\dots,(N_d)_{(d)}}=\sqrt{N_i}\Ket{(N_1)_{(1)},\dots,(N_i-1)_{(i)},\dots,(N_d)_{(d)}},\\ 
\fl b_i^\dagger\Ket{(N_1)_{(1)},\dots,(N_i)_{(i)},\dots,(N_d)_{(d)}}=\sqrt{N_i+1}\Ket{(N_1)_{(1)},\dots,(N_i+1)_{(i)},\dots,(N_d)_{(d)}}.
\end{eqnarray*}
Recall also the definitions of generators of the single-particle unitaries.
\begin{equation}
\fl\hat X_{i,j}:= b_i^\dagger b_j+b_j^\dagger b_i,\quad \hat Y_{i,j}:= i\left(-b_i^\dagger b_j+b_j^\dagger b_i\right), \quad
\hat Z_{i,j}:= b_i^\dagger b_i-b_j^\dagger b_j=\hat n_i-\hat n_j.
\end{equation}

We aim to study in detail the expression $\Bra{\Psi_{\epsilon,\delta}}\hat U^\dagger\hat W\hat U\Ket{\Psi_{\epsilon,\delta}}$ using expansion \Eref{uwu-generald}, i.e.
\begin{eqnarray}\label{goal}
\Bra{\Psi_{\epsilon,\delta}}\hat U^\dagger \hat W \hat U\Ket{\Psi_{\epsilon,\delta}}=\sum_{\gamma\leq \delta} \Big(\tilde\omega_{\gamma,\delta}\Bra{\Psi_{\epsilon,\delta}}\left(\hat R_\gamma\hat R_\delta+\hat R_\delta\hat R_\gamma\right)\Ket{\Psi_{\epsilon,\delta}}+\\\nonumber+ i\tilde\tau_{\gamma,\delta}\Bra{\Psi_{\epsilon,\delta}}\left(\hat R_\gamma\hat R_\delta-\hat R_\delta\hat R_\gamma\right)\Ket{\Psi_{\epsilon,\delta}}\Big).
\end{eqnarray}
In particular, we want to extract all terms which yield the constant $c_1$ from \Eref{bound-eps}, i.e. terms which are proportional to $\epsilon\mc{N}_\epsilon^2$ in \Eref{goal}. To this end, we focus on the expression $\left(b_l^\dagger b_k\pm b_k^\dagger b_l\right)\Ket{\Psi_{\epsilon,\sigma}}$, $k\leq l$, as its knowledge determines all the vectors $\hat R_\mu\Ket{\Psi_{\epsilon,\delta}}$ and thus all the expectation values in  \Eref{goal}.

If $k\neq 1$, then $\left(b_l^\dagger b_k\pm b_k^\dagger b_l\right)\Ket{N_{(1)}}=0$. Furthermore, if $k\neq 1$ and $k\neq i+1$ and $l\neq i+1$, $\left(b_l^\dagger b_k\pm b_k^\dagger b_l\right)\Ket{(N-2)_{(1)},2_{(i+1)}}=0$. Thus, the only nonzero contributions to $\left(b_l^\dagger b_k\pm b_k^\dagger b_l\right)\Ket{\Psi_{\epsilon,\sigma}}$ with $k\neq 1$ come from situations when either $k=i+1$ or $l=i+1$ in the \Eref{ansatz-app}. Thus, 
\begin{eqnarray}\label{action1}
\fl \left(b_l^\dagger b_k\pm b_k^\dagger b_l\right)\Ket{\Psi_{\epsilon,\sigma}}=\mc{N}_\epsilon\sigma\epsilon\left(\alpha_{k-1} b_l^\dagger b_k\Ket{(N-2)_1,2_{(k)}}\pm\alpha_{l-1} b_k^\dagger b_l\Ket{(N-2)_1,2_{(l)}}\right)\\\nonumber=\mc{N}_\epsilon\sigma\epsilon\sqrt{2}\left(\alpha_{k-1}\pm\alpha_{l-1}\right)\Ket{(N-2)_{(1)},1_{(k)},1_{(l)}}\quad {\mathrm{ if}} \quad k>1.
\end{eqnarray}
For $k=1$ and $2\leq l \leq d$, we have
\begin{eqnarray*}
\fl \left(b_l^\dagger b_1\pm b_1^\dagger b_l \right)\Ket{N_{(1)}}=\sqrt{N}\Ket{(N-1)_{(1)},1_{(l)}}, \\
\fl \left(b_l^\dagger b_1\pm b_1^\dagger b_l \right)\Ket{(N-2)_1,2_{(i+1)}}=\sqrt{N-2}\Ket{(N-3)_1,2_{(i+1)},1_{(l)}}\quad {\mathrm{ if}} \quad l\neq i+1, \\
\fl \left(b_l^\dagger b_1\pm b_1^\dagger b_l \right)\Ket{(N-2)_1,2_{(l)}}=\sqrt{3(N-2)}\Ket{(N-3)_1,3_{(l)}}\pm\sqrt{2(N-1)}\Ket{(N-1)_1,1_{(l)}}.
\end{eqnarray*}
Putting the above results together, we obtain
\begin{eqnarray}\label{action2}
\fl\left(b_l^\dagger b_1\pm b_1^\dagger b_l\right)\Ket{\Psi_{\epsilon,\sigma}}=\mc{N}_\epsilon\Big{(}\left(\sqrt{N}\pm \sigma\epsilon\alpha_{l-1}\sqrt{2(N-1)}\right)\Ket{(N-1)_1,1_{(l)}}+
\\\fl\nonumber+\sigma\epsilon\alpha_{l-1}\sqrt{3(N-2)}\Ket{(N-3)_1,3_{(l)}}+\sigma\epsilon\sqrt{N-2}\sum_{i=1,i\neq l-1}^{d-1}\alpha_{i}\Ket{(N-3)_1,2_{(i+1)},1_{(l)}}\Big{)}
\end{eqnarray}
For completeness, let us write down the action of $\hat Z_{k,l}$
\begin{eqnarray}\label{action3}
\left(\hat n_1-\hat n_l\right)\Ket{\Psi_{\epsilon,\sigma}}=\mc{N}_\epsilon\Big{(}N\Ket{N_{(1)}}+\sigma\epsilon\alpha_{l-1}(N-4)\Ket{(N-2)_{(1)},2_{(l)}}+\\ \nonumber +\sigma\epsilon(N-2)\sum_{i=1,i\neq l-1}^{d-1}\alpha_i\Ket{(N-2)_{(1)},2_{(i+1)}}\Big{)}, \\ \nonumber
\left(\hat n_k-\hat n_l\right)\Ket{\Psi_{\epsilon,\sigma}}=2\sigma\epsilon\left(\alpha_{k-1}-\alpha_{l-1}\right)\quad {\mathrm{ if}} \quad k>1.
\end{eqnarray}

By a visual inspection of \Eref{action1}, \Eref{action2}, and \Eref{action3} one can see that
\begin{enumerate}
\item $\Bra{\Psi_{\epsilon,\sigma}}\hat Z_{k,l}\hat X_{m,n}\Ket{\Psi_{\epsilon,\sigma}}=\Bra{\Psi_{\epsilon,\sigma}}\hat Z_{k,l}\hat Y_{m,n}\Ket{\Psi_{\epsilon,\sigma}}=0$ for any $k,l,m,n$ and the same holds for $\hat Y_{m,n}$.
\item Terms $\Bra{\Psi_{\epsilon,\sigma}}\hat Z_{k,l}\hat Z_{m,n}\Ket{\Psi_{\epsilon,\sigma}}$ do not produce any contribution proportional to $\epsilon \mc{N}_\epsilon^2$ for all $k,l,m,n$.
\item Terms $\Bra{\Psi_{\epsilon,\sigma}} \hat X_{k,l}\hat Y_{m,n}\Ket{\Psi_{\epsilon,\sigma}}$ could potentially yield some contribution proportional to $\epsilon \mc{N}_\epsilon^2$ if $k=m=1$ and $n=l$. However, the relevant term is $i\left(\sqrt{N}+ \sigma\epsilon\alpha_{l-1}\sqrt{2(N-1)}\right)\left(\sqrt{N}- \sigma\epsilon\alpha_{l-1}\sqrt{2(N-1)}\right)$ where the interesting terms cancel out in the end.
\item The only terms that yield expressions proportional to $\epsilon \mc{N}_\epsilon^2$ are $\Bra{\Psi_{\epsilon,\sigma}} \hat X_{1,l}^2\Ket{\Psi_{\epsilon,\sigma}}$ and $\Bra{\Psi_{\epsilon,\sigma}} \hat Y_{1,l}^2\Ket{\Psi_{\epsilon,\sigma}}$ for $1<l\leq d$. The relevant expressions read $\pm \mc{N}_\epsilon^2\left(\sqrt{N}\pm \sigma\epsilon\alpha_{l-1}\sqrt{2(N-1)}\right)^2$ which contribute via $\pm \sigma\epsilon\mc{N}_\epsilon^2\alpha_{l-1}2\sqrt{2N(N-1)}$. This leads directly to \Eref{c1d}.
\end{enumerate}

\subsection{The fermionic density functional}\label{appF}
Recall the ansatz singlet state
\begin{equation}\label{ansatz-f-app}
\fl \Ket{\Psi_{\epsilon,\sigma}^f}:=\mc{N}_\epsilon\left(1+\sigma\epsilon\sum_{i=1}^M\sum_{j=M+1}^d\alpha_{i,j}a_{j\downarrow}^\dagger a_{j\uparrow}^\dagger a_{i\uparrow} a_{i\downarrow}\right)\Ket{1\downarrow,1\uparrow,\dots,M\downarrow,M\uparrow},
\end{equation}
where $\mc{N}_\epsilon:=\frac{1}{\sqrt{1+\epsilon^2}}$, $\epsilon\geq 0$, $\sigma\in\{-1,+1\}$ and $\sum_{i=1}^M\sum_{j=M+1}^d\alpha_{i,j}^2=1$. We have also defined the single-particle operators of the $SU(d)$-algebra as 
\begin{equation}\label{generatorsd-f-app}
\fl \hat X_{i,j}^f:= f_{i,j}+f_{i,j}^\dagger,\quad \hat Y_{i,j}^f:= i\left(f_{i,j}-f_{i,j}^\dagger\right), \quad
\hat Z_{i,j}^f:= \left[f_{i,j}^\dagger,f_{i,j}\right]=\sum_{\sigma}\left(\hat n_{i\sigma}-\hat n_{j\sigma}\right)
\end{equation}
via the auxiliary "ladder" operators (in fact, these are the root operators of the $SU(d)$-algebra)
\[f_{i,j}:=\sum_{\sigma\in\{\uparrow,\downarrow\}}a_{j\sigma}^\dagger a_{i\sigma}, \quad 1\leq i\leq j\leq d.\]
We aim to evaluate the expression $\Bra{\Psi_{\epsilon,\delta}^f}\hat U^\dagger\hat W\hat U\Ket{\Psi_{\epsilon,\delta}^f}$ using the expansion (\ref{uwu-generald}). In particular, we want to find the constant $c_1^f$ in the RHS of \Eref{bound-eps-f}. 

One can verify in a straightforward way that the expressions of the form $\Bra{\Psi_{\epsilon,\sigma}^f}\hat Z_{k,l}^f\hat X_{m,n}^f\Ket{\Psi_{\epsilon,\sigma}^f}$ and $\Bra{\Psi_{\epsilon,\sigma}^f}\hat Z_{k,l}^f\hat Y_{m,n}^f\Ket{\Psi_{\epsilon,\sigma}^f}$ in \Eref{uwu-generald} vanish thanks to the special form of the ansatz state. Moreover, the expressions $\Bra{\Psi_{\epsilon,\sigma}}\hat Z_{k,l}^f\hat Z_{m,n}^f\Ket{\Psi_{\epsilon,\sigma}}$ give only terms proportional to $\mc{N}_\epsilon^2$ or to $\epsilon^2\mc{N}_\epsilon^2$, thus they do not contribute to the $c_1^f$.

Thus, the only expressions which may give terms proportional to $\epsilon\mc{N}_\epsilon^2$ in the \Eref{uwu-generald} are of the forms $\Bra{\Psi_{\epsilon,\sigma}^f} \hat X_{k,l}^f\hat X_{m,n}^f\Ket{\Psi_{\epsilon,\sigma}^f}$, $\Bra{\Psi_{\epsilon,\sigma}^f} \hat X_{k,l}^f\hat Y_{m,n}^f\Ket{\Psi_{\epsilon,\sigma}^f}$, $\Bra{\Psi_{\epsilon,\sigma}^f} \hat Y_{k,l}^f\hat Y_{m,n}^f\Ket{\Psi_{\epsilon,\sigma}^f}$. In order to evaluate these expressions, it is enough to find the vectors $\left(f_{k,l}\pm f_{k,l}^\dagger\right)\Ket{\Psi_{\epsilon,\sigma}^f}$ for $k<l$.

\begin{itemize}
\item Assume that $k>M$. Then, $f_{k,l}\Ket{1\downarrow,1\uparrow,\dots,M\downarrow,M\uparrow}=0$ and $f_{k,l}^\dagger\Ket{1\downarrow,1\uparrow,\dots,M\downarrow,M\uparrow}=0$. We also have
\begin{eqnarray*}\left(f_{k,l}\pm f_{k,l}^\dagger\right)a_{j\downarrow}^\dagger a_{j\uparrow}^\dagger a_{i\uparrow} a_{i\downarrow}\Ket{1\downarrow,1\uparrow,\dots,M\downarrow,M\uparrow}= \\
\fl=\delta_{k,j}\left(\Ket{1\downarrow,1\uparrow,\dots,M\downarrow,M\uparrow,j\downarrow,l\uparrow}-\Ket{1\downarrow,1\uparrow,\dots,M\downarrow,M\uparrow,j\uparrow,l\downarrow}\right)+ \\
\fl\pm\delta_{l,j}\left(\Ket{1\downarrow,1\uparrow,\dots,M\downarrow,M\uparrow,k\downarrow,j\uparrow}-\Ket{1\downarrow,1\uparrow,\dots,M\downarrow,M\uparrow,k\uparrow,j\downarrow}\right).
\end{eqnarray*}
\item Assume that $l\leq M$. Then, $f_{k,l}\Ket{1\downarrow,1\uparrow,\dots,M\downarrow,M\uparrow}=0$ and $f_{k,l}^\dagger\Ket{1\downarrow,1\uparrow,\dots,M\downarrow,M\uparrow}=0$. We also have
\small{
\begin{eqnarray*}
\left(f_{k,l}\pm f_{k,l}^\dagger\right)a_{j\downarrow}^\dagger a_{j\uparrow}^\dagger a_{i\uparrow} a_{i\downarrow}\Ket{1\downarrow,1\uparrow,\dots,M\downarrow,M\uparrow}= \\
\fl =\delta_{l,i}\big{(}\Ket{1\downarrow,1\uparrow,\dots,k\downarrow,\dots, i\uparrow\dots,M\downarrow,M\uparrow}-\Ket{1\downarrow,1\uparrow,\dots,k\uparrow,\dots, i\downarrow,\dots,M\downarrow,M\uparrow}\big{)}+ \\
\fl\pm\delta_{k,i}\big{(}\Ket{1\downarrow,1\uparrow,\dots,i\downarrow,\dots, l\uparrow\dots,M\downarrow,M\uparrow}-\Ket{1\downarrow,1\uparrow,\dots,i\uparrow,\dots, l\downarrow,\dots,M\downarrow,M\uparrow}\big{)}.
\end{eqnarray*}
}
\item Assume that $k\leq M$ and $l>M$. Then, $f_{k,l}^\dagger\Ket{1\downarrow,1\uparrow,\dots,M\downarrow,M\uparrow}=0$ and
\begin{eqnarray*}
\fl f_{k,l}\Ket{1\downarrow,1\uparrow,\dots,M\downarrow,M\uparrow}=\Ket{1\downarrow,1\uparrow,\dots,k\downarrow,\dots,M\downarrow,M\uparrow,l\uparrow}+ \\ -\Ket{1\downarrow,1\uparrow,\dots,k\uparrow,\dots,M\downarrow,M\uparrow,l\downarrow}.
\end{eqnarray*}
We also have
\begin{eqnarray*}
\fl f_{k,l}^\dagger a_{j\downarrow}^\dagger a_{j\uparrow}^\dagger a_{i\uparrow} a_{i\downarrow}\Ket{1\downarrow,1\uparrow,\dots,M\downarrow,M\uparrow}=\delta_{i,k}\delta_{j,l}\Big{(}\Ket{1\downarrow,1\uparrow,\dots,i\downarrow,\dots,M\downarrow,M\uparrow,j\uparrow}+ \\ -\Ket{1\downarrow,1\uparrow,\dots,i\uparrow,\dots,M\downarrow,M\uparrow,j\downarrow}\Big{)}, \\
\fl f_{k,l} a_{j\downarrow}^\dagger a_{j\uparrow}^\dagger a_{i\uparrow} a_{i\downarrow}\Ket{1\downarrow,1\uparrow,\dots,M\downarrow,M\uparrow}=(1-\delta_{i,k})(1-\delta_{j,l}) a_{j\downarrow}^\dagger a_{j\uparrow}^\dagger a_{i\uparrow} a_{i\downarrow}\times \\
\fl \times \Big{(}\Ket{1\downarrow,1\uparrow,\dots,k\downarrow,\dots,M\downarrow,M\uparrow,l\uparrow}-\Ket{1\downarrow,1\uparrow,\dots,k\uparrow,\dots,M\downarrow,M\uparrow,l\downarrow}\Big{)}.
\end{eqnarray*}
\end{itemize}
Thus, by the same reasoning as the one presented in \ref{appB} the only contributions to $\epsilon\mc{N}_\epsilon^2$ in the \Eref{uwu-generald} come from the expectation values  $\Bra{\Psi_{\epsilon,\sigma}^f} \left(\hat X_{k,l}\right)^2\Ket{\Psi_{\epsilon,\sigma}^f}$ and $\Bra{\Psi_{\epsilon,\sigma}^f} \left(\hat Y_{k,l}\right)^2\Ket{\Psi_{\epsilon,\sigma}^f}$ when $k\leq M$ and $l>M$. Each of such contributions comes from the expressions $2\mc{N}_\epsilon^2\left(1\pm\alpha_{k,l}\epsilon\sigma\right)^2$.

\end{document}